\documentclass{aa} %[a4paper,11pt, twoside ,openright]{aa}
\usepackage{graphicx}
\usepackage{txfonts}
\usepackage{lscape}
\usepackage{xcolor}
\usepackage{rotating}
\usepackage{graphicx}

\begin{document}

\title{Stellar Population Astrophysics (SPA) with TNG\thanks{Based on observations made with the Italian Telescopio Nazionale Galileo (TNG) operated on the island of La Palma by the Fundación Galileo Galilei of the INAF (Istituto Nazionale di Astrofisica) at the Observatorio del Roque de los Muchachos. This study is part of the Large Program titled SPA -- Stellar Population Astrophysics: a detailed, age-resolved chemical study of the Milky Way disc (PI: L. Origlia), granted observing time with HARPS-N and GIANO-B echelle spectrographs at the TNG.}}
\subtitle{The old open clusters Collinder~350, Gulliver~51, NGC~7044, and Ruprecht~171}
\author{G. Casali\inst{1,2},
L. Magrini\inst{2},
A. Frasca\inst{4},
A. Bragaglia\inst{3}, 
G. Catanzaro\inst{4},
V. D'Orazi\inst{5},
R. Sordo\inst{5},
E. Carretta\inst{3}, 
L. Origlia\inst{3}, 
G. Andreuzzi\inst{6,7},
X. Fu\inst{8},
A. Vallenari\inst{5}
%S. Lucatello\inst{5}
}
\institute{Dipartimento di Fisica e Astronomia, Universit\`a degli Studi di Firenze, via G. Sansone 1, 50019 Sesto Fiorentino (Firenze), Italy  \email{giada.casali@inaf.it} 
\and 
INAF-Osservatorio Astrofisico di Arcetri, Largo E. Fermi, 5, I-50125 Firenze, Italy 
\and
INAF-Osservatorio di Astrofisica e Scienza dello Spazio di Bologna, via P. Gobetti 93/3, 40129, Bologna, Italy
\and
INAF-Osservatorio Astrofisico di Catania, via S. Sofia 78, 95123, Catania, Italy
\and
INAF-Osservatorio Astronomico di Padova, vicolo Osservatorio 5, 35122, Padova, Italy
\and
Fundaci\'{o}n Galileo Galilei - INAF, Rambla Jos\'{e} Ana Fern\'{a}ndez P\'{e}rez 7, 38712 Bre\~na Baja, Tenerife, Spain
\and
INAF-Osservatorio Astronomico di Roma, Via Frascati 33, 00078 Monte Porzio Catone, Italy
\and
The Kavli Institute for Astronomy and Astrophysics at Peking University, 100871 Beijing, China
}

 \abstract
   {Open clusters are excellent tracers of the chemical evolution of the Galactic disc. The spatial distribution of their elemental abundances, through the analysis of high-quality and high-resolution spectra, provides insight into the chemical evolution and mechanisms of element nucleosynthesis in regions characterised by different conditions (e.g. star formation efficiency and metallicity). }
   {In the framework of the Stellar Population Astrophysics (SPA) project, we present new observations and  spectral analysis of four sparsely studied open clusters located in the solar neighbourhood, namely \object{Collinder~350}, \object{Gulliver~51}, \object{NGC~7044,} and \object{Ruprecht~171}.  }
   {We exploit the HARPS-N spectrograph at the TNG telescope to acquire high-resolution optical spectra for 15 member stars of four clusters. We derive stellar parameters ($T_{\rm eff}$, $\log~g$, [Fe/H] and $\xi$) using both the equivalent width (EW) analysis and the spectral fitting technique. We compute elemental abundances for light, $\alpha$-, iron-peak, and n-capture elements using the EW measurement approach. We investigate the origin of the correlation between metallicity and stellar parameters derived with the EW method for the coolest stars of the sample ($T_{\rm eff}< 4300$ K). The correlation is likely due to the challenging continuum setting and to a general inaccuracy of model atmospheres used to reproduce the conditions of very cool giant stars. }
   {We locate the properties of our clusters in the radial distributions of metallicity and abundance ratios, comparing our results with clusters from the Gaia-ESO and APOGEE surveys. We present the [X/Fe]-[Fe/H] and [X/Fe]-$R_{\rm GC}$ trends for elements in common between the two surveys. Finally, we derive the C and Li abundances as a function of the evolutionary phase and compare them with theoretical models.
   }
   {The SPA survey, with its high-resolution spectra, allows us to fully characterise the chemistry of nearby clusters. With a single set of spectra, we provide chemical abundances for a variety of chemical elements, which are comparable to those obtained in two of the largest surveys combined.  
   The metallicities and abundance ratios of our clusters fit very well in the radial  distributions defined by the recent literature, reinforcing the importance of star clusters to outline the spatial distribution of  abundances in our Galaxy. Moreover, the abundances of C and Li, modified by stellar evolution during the giant phase, agree with evolutionary prescriptions (rotation-induced mixing) for their masses and metallicities.
   }

\authorrunning{Casali, G. et al.}
\titlerunning{SPA-OCs}

\maketitle

\section{Introduction}
Star clusters are among the most versatile astronomical objects. Within our Galaxy, they play a key role in the study of both stellar and Galactic evolution: they allow us to study the formation and evolution of stars \citep[e.g.][]{krause20, kumholz20}, the dynamics of stellar systems \citep[e.g.][]{sacco17a,kuhn19, piatti19}, and they provide robust constraints on the formation timescales and the chemical and dynamical history of the Milky Way \citep[e.g.][]{friel02,bt06,magrini09, reddy16, jaco16, spina17, occaso19, zhong20, donor20, chen20}.

Large spectroscopic surveys, such as for example Gaia-ESO \citep{Gil}, GALAH \citep{gala}, and APOGEE \citep{majewski17}, and future survey-dedicated spectrographs, such as WEAVE \citep{dalton12} and 4MOST \citep{4most}, make use of multi-object spectroscopy at medium-high resolution to characterise the kinematics and global chemical properties of the different Galactic stellar components (i.e. disc, bulge, and halo) with high-statistical significance. Their observations complement the data from the {\em Gaia} mission \citep{perryman01,gaia1, gaia2}, which in its second data release \citep[{\em Gaia} DR2,][]{gaiadr2} has provided positions, parallaxes, proper motions, and photometry in three bands ($G$, $BP$, $RP$) for more than 1.3 billion sources.

A large number of star clusters are included among the targets of the large spectroscopic surveys. 
In particular, the Gaia-ESO survey provides data for 81 open clusters, in most cases with more than 100 member stars; APOGEE  presently has spectra for 128 open clusters \citep{donor20}, but generally with only a few stars per cluster; WEAVE, which will begin soon, will target about 300 open clusters. 
The observations of large spectroscopic surveys, together with {\em Gaia} data, have improved our understanding of Galactic chemical evolution \citep[e.g.][]{magrini17, magrini18, donor20}, the Milky Way structure \citep[e.g.][]{meingast19, cantat20, castro20, anders20} and the cluster formation and disruption processes \citep[e.g.][]{sacco17,bravi, piatti19}. 
In addition, {\em Gaia} has enabled the discovery and characterisation of a large number of clusters in the solar neighbourhood and beyond \citep{cantat18a,cantat18b,cantat20,sim19,lp19,castro18,castro20}, and at the same time has allowed some candidate clusters to be discarded  \citep[see e.g.][for discussion]{kos18,cantat20a}, and several extended structures to be identified, such as strings 
and filaments, which are often connected to known clusters \citep[e.g.][]{kc19}, streams \citep[e.g.][]{meingast19}, extended halos \citep[e.g in M67 by][]{carrera19}, and tidal tails \citep[e.g. in Praesepe by][]{rs19}. {\em Gaia} data, in combination with spectral information on kinematics (from radial velocities) and abundances, are driving a revolution in the study of open clusters and, in general, of the whole Milky Way.

In this framework, the aims of the Stellar Population Astrophysics (SPA) project, an ongoing Large Programme running on the 3.6 m Telescopio Nazionale Galileo (TNG) at the Roque de los Muchachos Observatory (La Palma, Spain), are to contribute to our understanding of the star formation and chemical enrichment history of our Galaxy by providing high-resolution spectra of a sample of stars in the Solar neighbourhood.  
The SPA project is obtaining high-resolution spectra of approximately 500 stars near to the Sun, covering a wide range of ages and properties \citep[see][for a general description]{origlia19}, such as Cepheids
and stars in both young and old open clusters of spectral type from A to K.
These stars are observed in the optical and near-infrared (NIR) bands at high spectral resolution using GIARPS, a combination of HARPS-N and GIANO-B spectrographs. 
The aim of SPA is to obtain a large variety of elemental abundances in order to seek possible global trends, such as for example [X/Fe]--age relations. Chemical characterisation combined with the kinematic counterpart from {\em Gaia} and other surveys will provide a framework for a comprehensive chemo-dynamical modelling of disc formation and evolution in the Solar vicinity. 
The present study is part of a series of papers dedicated to the results of the SPA project. So far, the series includes a paper on the red supergiants in Alicante 7 and Alicante 10 \citep{origlia19}, and studies of the young open clusters ASCC~123 \citep{frasca19} and Praesepe \citep{dorazi20}. 
Part of our goal is to gain kinematic and chemical information on open clusters that have never been studied before or for which very little spectroscopic data is available. This is the case of the four clusters presented here, Collinder~350, Gulliver~51, NGC~7044, and Ruprecht~171. One of them, Gulliver~51, is a new cluster discovered using {\em Gaia} DR2 data \citep{cantat18a}, while   only one star was observed spectroscopically in Collinder~350, and NGC~7044 has only been studied with low-resolution spectra.

The paper is structured as follows: in Sect.~\ref{clusters} we present our cluster sample, and in Sect~\ref{obs} we show the spectral data sample collected with the spectrograph HARPS-N at the TNG telescope during the SPA observing campaign. In Sect.~\ref{sec:photoinput} we present the photometric parameters for our star sample, and in Sect.~\ref{analysis}, we describe our spectral analysis using both the equivalent width (EW) analysis and the spectral fitting. In Sect.~\ref{discussion} we discuss the discrepancy in metallicity between cool and warm member stars of the same cluster, and in Sects.~\ref{abu} and~\ref{res}, we review the chemical abundance ratios of our clusters in the context of larger data samples which define the abundance gradients [X/Fe] versus [Fe/H] and [$\alpha$/Fe] versus [Fe/H] and the radial metallicity gradient. We present the C and Li abundances as a function of $\log~g$ compared with the evolutionary predictions. Finally, in Sect.~\ref{conclusions}, we summarise our results and give our conclusions.

\section{The cluster sample}
\label{clusters}

The SPA project is observing a large sample of nearby star clusters, allowing complete characterisation of the open clusters located in the Solar neighbourhood.
Most of these clusters are located within 1.5-2 kpc of the Sun in order to match the zone where {\em Gaia} data reach their highest precision. We need to obtain, in a reasonable amount of time, sufficiently high $S/N$  to permit precise abundance determination from spectra  at the very high resolution of GIARPS. This means that our magnitude limit is about $G=12-12.5$ mag. The targets are main sequence stars in the closer and younger clusters and giant stars in the older ones.
 In general, we try to target stars on the red clump, because they are suitable for precise analysis and constitute a homogeneous sample with many other previous projects \citep[e.g.][for the BOCCE and OCCASO projects, respectively]{bocce,occaso}. In a few cases, for interesting  clusters that are more distant and/or extincted, only stars on the brighter part of the red giant branch can be observed.

In the present work, we discuss the analysis of four open clusters that are deemed important to characterise the nearby regions of our Galaxy because of their location and age, but for which very little spectroscopic data are available. 
Their ages range between 0.3 and 3 Gyr and they are located from $\sim$300~pc to about 3300~pc from the Sun at different Galactocentric distances and altitudes above the Galactic plane.

Collinder~350 was listed for the first time in the catalogue of \citet{collinder31}, but the first study of its properties was presented by \citet{kharchenko05}. The cluster was also studied using {\em Gaia} DR1/TGAS data by \citet{cantat18b} and \citet[][who derived an age of 1 Gyr]{yen2018}. 
\citet{pakhomov09} obtained high-resolution spectroscopy of one star, indicated as HD161587 (our Cr350\_1), and derived a metallicity of [Fe/H]$=+0.11 \pm 0.06$ dex %Teff/logg/csi=4270/1.47/1.66
together with the abundances of many species. \citet{blanco15} and \citet{blanco18}, also from one single star (at resolution  R=80~000, with the NARVAL spectrograph), derived a lower value for the metallicity, [Fe/H]$=-0.10 \pm 0.01$ (or 0, depending on normalisation) and 0.03 dex, respectively. On the other hand, there are no high-resolution spectroscopic observations of the other three clusters.

NGC~7044 has been studied using photometry several times in the past \citep[see][]{kaluzny89,aparicio93,sagar98}, with a general consensus on its age of around 1.5 Gyr, and a high reddening. An estimate of the metallicity and radial velocity of ten member stars in NGC~7044 is provided by \citet{Warren09} using low-resolution spectroscopy and the IR Ca\,{\sc ii} triplet (CaT) technique. These authors derived a mean metallicity [Fe/H]$=-0.16 \pm 0.09$ dex and a mean radial velocity RV$=-50.56 \pm 2.18$ km s$^{-1}$. 

Even less information is available for Ruprecht~171: after the classification by \citet{ruprecht66}, the cluster was studied by \citet{Tadross} who derived  an
age 3.2 Gyr, a reddening $E(B-V)=0.12$ mag, and a distance $d=1140$ pc using the isochrone fitting of the NIR $JHK_{s}$ photometric data. 

Finally, Gulliver~51 was only recently discovered by \citet{cantat18a}, who reported the serendipitous discovery of 60 candidate clusters based on {\em Gaia} DR2 data, which were named `Gulliver'. They were identified as groups of stars with coherent proper motions and parallaxes, and  a more concentrated distribution on the sky  than the field population.  An age of 0.8 Gyr,  reddening $E(B-V)=0.375$ mag, and a distance of 1330 pc were attributed to Gulliver~51
by \citet{monteiro19} based on automated isochrone fitting to {\em Gaia} DR2 colour-magnitude diagrams (CMDs).

\begin{figure*}[h]
\centering
\hspace{-0.5cm}
\includegraphics[scale=0.5]{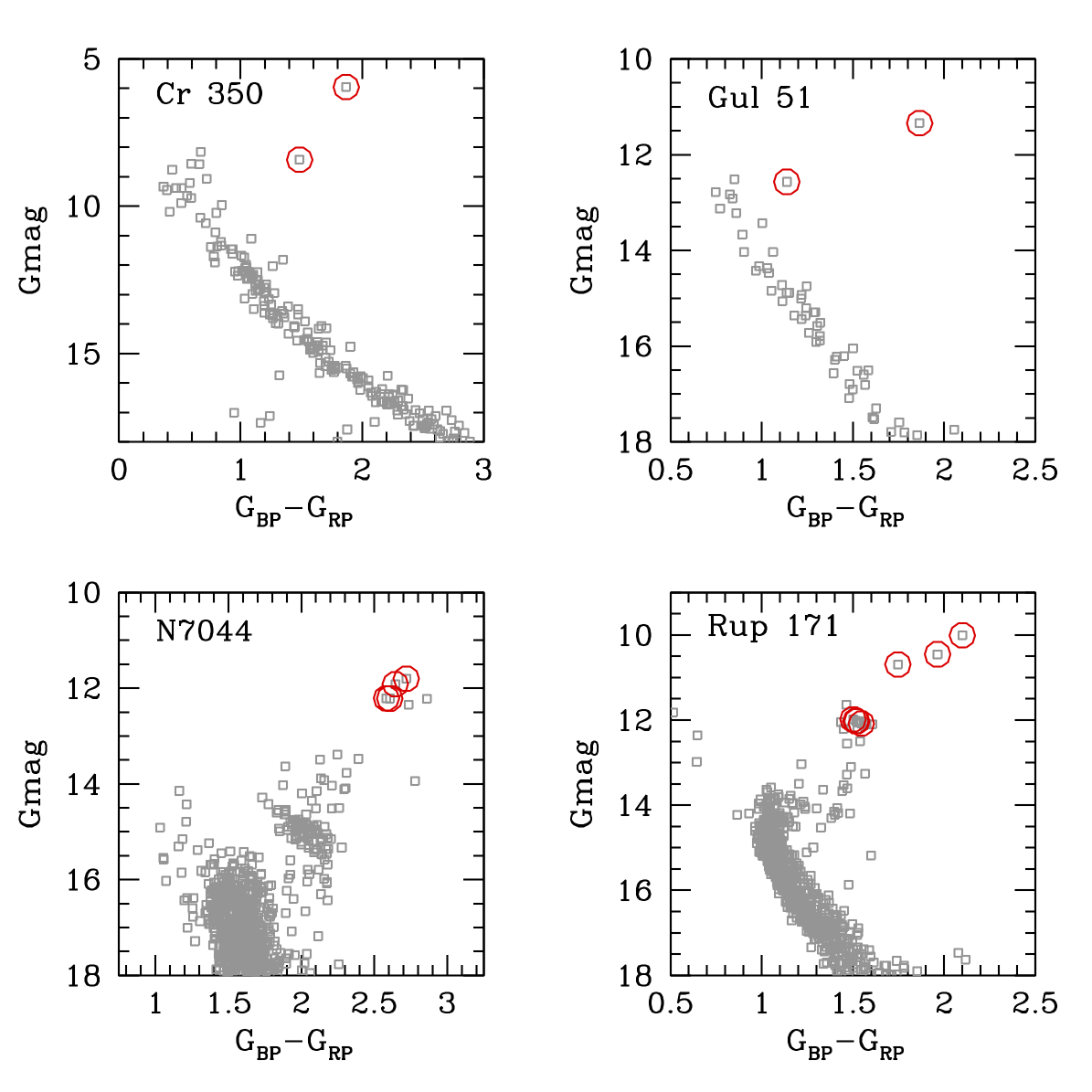}
\caption{Colour-magnitude diagrams with {\em Gaia} DR2 photometric data (G$_{\rm mag}$ vs. G$_{\rm BP}-$G$_{\rm RP}$) of the four clusters. The red circles indicate the member stars observed by SPA project.}
\label{fig:cmd}
\end{figure*}

For the first three clusters, distance, reddening, age, and proper motions are reported in the catalogue of \citet{khar13}. All have updated values by \citet{cantat18a,cantat20} using {\em Gaia} DR2 data. Table~\ref{tab:infoclusters} gives the coordinates, proper motions ($\mu_{\alpha}$,  $\mu_{\delta}$), parallax, distance ($d$), Galactocentric distance ($R_{\rm GC}$), altitudes above the Galactic plane ($z$), extinction ($A_{V}$), and age of the clusters by \citet{cantat20}. Figure~\ref{fig:cmd} shows the CMDs of the four clusters, in which the stars observed by the SPA project are highlighted. 

This is the first paper dedicated exclusively to these four clusters. In this work we provide, for the first time (except for Collinder~350, for which high-resolution spectroscopy of a single star is available), results based on high-resolution spectroscopy for a number of candidate members in each cluster. Our aim is to perform a full characterisation in terms of atmospheric parameters and chemical abundances.

\begin{table*}[ht]
\caption{Parameters of the open clusters sample from \citet{cantat20}.}
\begin{center}
%\fontsize{8pt}{8pt}
\setlength{\tabcolsep}{5pt}
\selectfont
\small{
\begin{tabular}{lcccccccccc}
\hline
\hline
Cluster        &   R.A.        & Dec.           &  $\mu_{\alpha}$   &  $\mu_{\delta}$   &  parallax        & log(Age) & $A_{V}$ &   $d$ & $R_{\rm GC}$ & $z$\\ 
               &  (J2000)      &                & (mas~yr$^{-1}$)   &  (mas~yr$^{-1}$)  &  (mas)           &  (yr)      & (mag)  &   (pc)  & (kpc) & (pc)\\
\hline
Collinder~350  &  17:48:14.26  &   +01:20:25.42 & $-4.965\pm 0.387$ & $-0.019\pm 0.243$ & $2.708\pm 0.129$ &  8.77    & 0.52 &  371 & 8.02 &94\\  
Gulliver~51    &  02:01:20.40  &   +63:48:03.60 & $-4.892\pm 0.097$ & $-0.149\pm 0.079$ & $0.647\pm 0.026$ &  8.56    & 1.42 & 1536 & 9.41 & 52\\
NGC~7044       &  21:13:08.16  &   +42:29:38.40 & $-4.978\pm 0.142$ & $-5.526\pm 0.151$ & $0.273\pm 0.078$ &  9.22    & 1.78 & 3252 & 8.73 & $-235$\\
Ruprecht~171   &  18:32:02.88  & $-$16:03:43.20 & $+7.677\pm 0.187$ & $+1.091\pm 0.165$ & $0.620\pm 0.066$ &  9.44    & 0.68 & 1522      & 9.41 & 52\\
 \hline
\end{tabular}
}
%\tablefoot{Each value is taken from \citet{cantat20}, except for (1) \citet{blanco18} and (2) \citet{Warren}.}
%\tablefoot{Notes: $^{a}$ \citet{khar13}, $^{b}$ \citet{dias12}, $^{c}$ \citet{blanco18}, $^{d}$ \citet{Warren}, $^{e}$ \citet{cantat20}. (1) \citet{yen2018}; (2) \citet{cantat18a}} 
%0.01 \citet[phtometrically]{Tadross}
\end{center}
\label{tab:infoclusters}
\end{table*}

\section{Observations and data sample}
\label{obs}
We select high-probability member stars among the red giant branch (RGB) and red clump (RC) stars. The membership probability was taken from \citet{cantat18a} who used {\em Gaia} DR2 proper motions and parallaxes.
The observations of these four clusters were conducted from 18 to 22 August 2018 and from 10 to 15 August 2019 with GIARPS using both the optical echelle spectrograph HARPS-N \citep[R $\sim$ 115~000, spectral range $= 0.39-0.68~\mu m$,][]{cosentino14} and the NIR spectrograph GIANO-B \citep[R $\sim$ 50~000, spectral range $=0.97-2.45~\mu m$,][]{oliva06}. 
The spectra were acquired with total exposure times ranging from 600 to 7200 seconds depending on the star brightness, in order to reach an S/N per pixel at red wavelengths of $S/N > 30$. Exposure times longer than 1800 seconds were usually split into two or three sub-exposures to reduce the contamination of cosmic rays and to avoid saturation.
The stars analysed in the present work are shown in Table~\ref{tab:stars} with their coordinates, {\it Gaia} magnitudes, parallax, radial velocity (RV), exposure time, and S/N.
The RV was measured by cross-correlating the target spectrum with a template, which was chosen as the 
synthetic BT-Settl spectrum  \citep{Allard11} with $T_{\rm eff}$ and $\log g$ closer to the target.
Very broad lines, such as \ion{Na}{i}\,D$_2$ and Balmer lines, as well as strong telluric features were excluded from the  cross-correlation function CCF analysis. 
For this task we used ad hoc software developed by us in the IDL environment.
The CCF peak was fitted with a Gaussian to evaluate its centroid and full width at half maximum (FWHM). The RV error
was estimated by the fitting procedure accounting for the CCF noise far from the peak. Here we only make use of HARPS-N spectra, which are better suited for our analysis methods (see following section). GIANO spectra will be presented in a forthcoming paper. 

\section{Photometric parameters}
\label{sec:photoinput}

In Table~\ref{tab:input}, we present the photometric parameters $T_{\rm eff,\, Gaia}$ and  $\log~g_{\rm \, Gaia}$ obtained from {\em Gaia} DR2 photometry and parallaxes, and the parameters obtained from the comparison with the best-fit isochrone ($T_{\rm eff,\, iso}$ and $\log~g_{\rm \, iso}$), projecting the {\it Gaia} colours and magnitudes on a set of PARSEC isochrones \citep{Bressan12}.  
Our aim is to use these parameters as an input to estimate the spectroscopic ones.
The photometric gravities from {\em Gaia} photometry are obtained using the following equation:
\begin{equation}
 \log(g_{\rm \, Gaia}) = \log(M/M_{\odot}) + 0.4 \cdot M_{bol} + 4 \cdot \log(T_{\rm eff,\, Gaia}) - 12.505,
 \end{equation}
where $M/M_{\odot}$ is the stellar mass (in solar mass units) obtained as the mass at the main sequence turn off (MSTO) of the isochrones at their literature cluster age; and $M_{bol}$ ($M_{bol} = - 2.5 \cdot \log(L/L_{\odot}) + 4.75 $) is the bolometric magnitude computed from the luminosity present in the {\em Gaia} DR2 catalogue \citep{Li2018} and corrected considering the average distance for each cluster. In parenthesis in Table~\ref{tab:input}, we give the $\log~g$ with $M_{bol}$ computed with the luminosity corrected considering the individual distances derived by \citet{Bailer-Jones18} and the mean extinction values of the clusters reported in Table~\ref{tab:infoclusters}; $T_{\rm eff,\, Gaia}$ is the photometric effective temperature from {\em Gaia} obtained with the calibration of giant stars by \citet{mb20} for the {\it Gaia} colour $BP-BR$. In first approximation, we consider solar metallicities for our clusters. As a test, we also re-compute the photometric effective temperatures using average metallicities from spectroscopic analysis, finding a negligible correction.  For the coolest stars of the sample (all stars of NGC~7044 and Rup171\_1), as their $BP-RP$ colours fall outside the calibration range of \citet{mb20}, we adopt the effective temperature of {\it Gaia} DR2 as $T_{\rm eff,\, Gaia}$, for which we are aware of the
inadequacy, especially at those very low temperatures and for cluster members \citep[see][]{andrae18,gaiadr2}. 
For clusters affected by high extinction and in crowded regions, the photometric parameters are indeed unreliable because of the spatial variations of the reddening, which is usually assumed to be constant,  and because of the more difficult extraction of the fluxes of the individual stars.   
The photometric gravities of the four observed stars of NGC~7044, the most distant and extincted cluster of our sample, range from 0.8 to 1.34 dex (obtained with the individual distances), which is unexpected, because they have very similar colours and magnitudes (see Fig.~\ref{fig:cmd}). Their parallaxes have percentage errors larger than 10\%, and they differ by more than 3$\sigma$ from the mean cluster parallax. In addition, the individual extinctions available in {\it Gaia} DR2 (albeit suffering from the same limitation as {\it Gaia} DR2 $T_{\rm eff}$) vary from star to star, and are not available for all the cluster members. Therefore, the photometric parameters of the observed stars in NGC~7044 have to be considered as just a starting point for the following spectroscopic analysis. 
In a similar way, the two stars observed in Gul~51  both  have high extinction in {\it Gaia} DR2, which varies from star to star ($A_{G}\sim 1$ mag for Gul~51\_1 and more than 2.5 mag for Gul~51\_2). We adopt the average cluster extinction as in \citet{cantat20}, however differential extinction can significantly affect photometric stellar parameters, with strong effects on the photometric $T_{\rm eff}$ and $\log~g$. 
In Ruprecht~171, we observed seven stars, four of them with very similar colours and magnitudes. This cluster is closer and less affected by extinction. The photometric parameters of the four hottest stars (Rup171\_4-5-6-7) are indeed very similar. Finally, Collinder~350 is close and not significantly affected by extinction. 

Regarding the stellar parameters  $T_{\rm eff, iso}$ and $\log~g_{\rm iso}$, we select the best isochrone for each cluster, starting from the cluster parameters of \citet{cantat20} and  fine-tuning them with our grid of isochrones.
The presence of high differential reddening produces, in some cases, differences between the two methods for estimating the photometric stellar parameters and confirms that they can only be used as priors for our spectroscopic analysis.

\begin{table*}[ht]
\caption{Observed candidate member stars for the four clusters.}
\begin{center}
\small{
\begin{tabular}{lccccccccc}
\hline
\hline
 Gaia DR2 ID &     ID        & R.A.    & Dec. & $G$ & $BP-RP$ & RV            & Exposure Time & S/N & parallax \\ 
             &               & (J2000) &      & (mag)  &    (mag)    & (km s$^{-1}$) & (s)           &     & (mas) \\
\hline
4372743213795720704  & Cr350\_1   & 266.603650 &  +1.044348     &  5.96 & 1.87 & $-14.57$ & 600   &  172 &2.910$\pm$0.061\\ 
4372572888274176768  & Cr350\_2   & 267.182595 &  +1.164203     &  8.42 & 1.48 & $-14.73$ & 1800  &  158 &2.770$\pm$0.074\\
517925575042048384   & Gul51\_1   &  30.326628 &  +63.79737     & 11.34 & 1.86 & $-57.36$ & 7200  &   68 &0.725$\pm$0.025\\ 
517953750028538240   & Gul51\_2   &  30.213987 &  +63.96327     & 12.57 & 1.14 & -- & 7200  &   34 & 0.645$\pm$0.030 \\ 
1969807040026523008  & NGC7044\_1 & 318.321873 &  +42.48457     & 11.80 & 2.72 & $-49.88$ & 7200  &   46 &0.336$\pm$0.034\\ 
1969807276235623552  & NGC7044\_2 & 318.330484 &  +42.50797     & 11.92 & 2.64 & $-49.15$ & 7200  &   36 &0.216$\pm$0.035\\ 
1969806073644788992  & NGC7044\_3 & 318.397775 &  +42.46081     & 12.21 & 2.58 & $-49.45$ & 7200  &   34 &0.280$\pm$0.032\\ 
1969800576086654592  & NGC7044\_4 & 318.256943 &  +42.40348     & 12.22 & 2.61 & $-49.44$ & 7200  &   31 &0.333$\pm$0.035\\ 
4103073693495483904  & Rup171\_1  & 277.989813 &$-$15.98095     & 10.01 & 2.10 & $5.55$ & 3600  &   53 &0.746$\pm$0.061\\ 
4102882309792631552  & Rup171\_2  & 278.022116 &$-$16.13376     & 10.45 & 1.96 & $5.32$ & 7200  &   66 &0.638$\pm$0.046\\ 
4103072765721906816  & Rup171\_3  & 278.033168 &$-$16.00760     & 10.69 & 1.75 & $6.84$ & 7200  &   85 &0.667$\pm$0.054\\ 
4103101073850814208  & Rup171\_4  & 278.054888 &$-$15.87125     & 12.02 & 1.52 & $5.46$ & 7200  &   53 &0.626$\pm$0.044\\ 
4102884023383492096  & Rup171\_5  & 278.030396 &$-$16.10391     & 12.03 & 1.52 & $6.03$ & 7200  &   45 &0.677$\pm$0.044\\ 
4103073418617487104  & Rup171\_6  & 278.013899 &$-$15.99949     & 12.03 & 1.52 & $5.81$ & 7200  &   38 &0.629$\pm$0.043\\ 
4103072933225072512  & Rup171\_8  & 277.945748 &$-$16.03430     & 12.06 & 1.53 & $6.28$ & 7200  &   36 &0.639$\pm$0.043\\ 
  %23473588+6833329 & King~11         & 356.899458 &  +68.55919     & 12.07 & 3.03  & \\

 \hline
\end{tabular}
}
%\tablefoot{}
\end{center}
\label{tab:stars}
\end{table*}

\begin{table}[ht]
\caption{Photometric stellar parameters from {\it Gaia} DR2 and PARSEC isochrones.}
\begin{center}
%\fontsize{6pt}{8pt}
\setlength{\tabcolsep}{5pt}
\begin{tabular}{lcccc}%cc}
\hline
\hline
 ID        & $T_{\rm eff, Gaia}$ & $\log~g_{\rm Gaia}$ & $T_{\rm eff, iso}$ & $\log~g_{\rm iso}$\\%& [Fe/H] & $\xi$ \\ 
          & (K) &   (dex)  & (K) &   (dex)  \\ %& (dex)  &    (km/s)    \\
\hline
 Cr350\_1   & 4030 & 1.24 (1.30)   &  4100 &  1.35 \\%&  $-$0.10 & 1.85  \\
 Cr350\_2   & 4620 & 2.51 (2.54)   &  4880 &  2.71 \\ %&  $-$0.10 & 1.49  \\
 Gul51\_1   & 4453 & 1.90 (2.02)   &  4630 &  2.26 \\%&  0.0     & 1.64 \\
 Gul51\_2   & 6122*  & 2.93 (2.92) &  6610 &  3.11  \\
 NGC7044\_1 & 3825** & 0.89 (1.04) &  3850 &  0.98 \\%&  $-$0.16 & 1.93  \\
 NGC7044\_2 & 3898** & 0.98 (0.82) &  3900 &  1.06 \\%&  $-$0.16 & 1.91  \\
 NGC7044\_3 & 3929** & 1.13 (1.14) &  4000 &  1.23 \\%&  $-$0.16 & 1.86  \\
 NGC7044\_4 & 4021** & 1.20 (1.34) &  4000 &  1.23 \\%&  $-$0.16 & 1.86  \\
 Rup171\_1  & 3863** & 1.25 (1.39) &  3930 &  1.25 \\%&  0.30  & 1.99  \\
 Rup171\_2  & 3940 & 1.52 (1.53)   &  4090 &  1.54 \\%&  0.30  & 1.90  \\
 Rup171\_3  & 4220 & 1.68 (1.72)   &  4300 &  1.76 \\%&  0.30  & 1.85  \\
 Rup171\_4  & 4600 & 2.42 (2.41)   &  4750 &  2.52 \\%&  0.30  & 1.65  \\
 Rup171\_5  & 4600 & 2.43 (2.48)   &  4750 &  2.52 \\%\\ % 2.64 \\ %&  4400 &  2.64 &  0.30  & 1.46  \\
 Rup171\_6  & 4600 & 2.44 (2.43)   &  4750 &  2.52 \\%\\ %2.58&  4420 &  2.60 &  0.30  & 1.49  \\
 Rup171\_8  & 4600 & 2.38 (2.39)   &  4750 &  2.52 \\%\\ %&2.55  4310 &  2.56 &  0.30  & 1.48  \\
 \hline
\end{tabular}
\tablefoot{(*) $T_{\rm eff}$ from the calibration of dwarf stars in \citet{mb20}; (**) $T_{\rm eff}$ from {\it Gaia} DR2. }

\end{center}
\label{tab:input}
\end{table}

\section{Spectral analysis}
\label{analysis}
We follow two different approaches to analyse our sample stars. The first one is a spectral analysis using the EWs with Fast
Automatic MOOG Analysis  \citep[FAMA,][]{magrini2013}, while the second one is a spectral fitting with ROTFIT \citep{frasca06, frasca19}.

\subsection{Equivalent width analysis with Fast
Automatic MOOG Analysis}

We perform a spectral analysis based on the EWs to determine the atmospheric parameters and abundances of our sample stars. We measure the EWs of the spectral absorption lines with the {\sc Daospec} \citep{daospec} tool \citep[in the form of DOOp - \textsc{DAOSPEC} Output Optimiser pipeline, an automatic wrapper;][]{cantat14}. 
We use the master list of atomic transitions that was prepared for the analysis of the stellar spectra for the Gaia-ESO survey \citep{heiter15}. 
This line list includes quality flags, such as `Y' (yes), `N' (no), and `U' (undetermined). These flags are assigned on the basis of the quality of the line profiles  (at the spectral resolution of about 47000) and the accuracy of the log~$gf$ derived from the comparison of synthetic spectra with a spectrum of the Sun and Arcturus. If the profile of a given line is unblended in both the Sun and Arcturus and its log~$gf$ value is well determined, the flag will be `Y/Y', and `N/N' otherwise. In our analysis, we consider all lines, except those with `N' for the log~$gf$ values.
At high-spectral resolution, the line profile might differ from the Gaussian profile, especially for the strongest lines. However, this effect is usually negligible at least for EWs~$<$~100$-$120 m\AA, which corresponds to the range we consider for the analysis, as discussed in \citet{spina20}, who tested the effect of the use of a Voigt profiles in the measurement of the EWs of strong lines for HARPS spectra. The two measurements, that is those made with a Gaussian and those with a Voigt profile, are consistent, and the uncertainty due to the choice of the profile is negligible with respect to the uncertainty related to continuum placement.

%It includes 862 lines of Fe I, 42 of Fe II and 780 for the other elements, detectable in the spectral range (4200-9200 $\AA$). 
We determine the atmospheric parameters with MOOG in the automatic form \citep[FAMA,][]{magrini2013}. FAMA uses MOOG in its 2014 version \citep{sneden12} and MARCS model atmospheres \citep{Gustafsson08} to determine stellar parameters and to calculate abundances.
It iteratively searches for the three equilibria: the ionisation equilibrium, excitation equilibrium, and minimisation of the trend between the reduced EW $\log (\rm EW/\lambda )$ and the iron abundance. 
The effective temperature $T_{\rm eff}$ is obtained by minimising the slope between iron abundance and the excitation potential; the surface gravity $\log~g$ is obtained by assuming the ionisation equilibrium condition between Fe\,{\sc i} and Fe\,{\sc ii}; and the microturbulence $\xi$ is obtained by minimising the trend between iron abundance and the reduced EWs. 
To avoid saturated and overly weak lines,  in our analysis we only consider lines with 20 m$\AA$ < EW < 120 m$\AA$ for iron, and 5m$\AA$ < EW < 120 m$\AA$ for other elements.

The evaluation of the uncertainties on the final stellar parameters is done when the slope of the trend between the Fe abundance and reduced EWs, and the difference between Fe\,{\sc i} and Fe\,{\sc ii} abundances, are not completely minimised. FAMA uses the dispersion of the abundances to derive the errors on the atmospheric parameters: $T_{\rm eff}$ error is the ratio between $\sigma_{\rm Fe\,\textsc{i}}$, the dispersion around the mean of $\log n(\rm Fe\,\textsc{i})$, and the range of excitation potential EP; $\log~g$ error is $\sqrt{\sigma_{\rm Fe\,\textsc{i}}^2 + \sigma_{\rm Fe\,\textsc{ii}}^2}$; and $\xi$ error is the ratio between $\sigma_{\rm Fe\,\textsc{i}}$ and the range of reduced EW. There are two types of errors in metallicity: {\em (i)} the statistical uncertainty due to the random errors in the EW measurements and to uncertainties on the atomic parameters; and {\em (ii)} the errors on the abundances generated by the uncertainties in the determination of the atmospheric parameters. Both are indicated in Table~\ref{tab:clusters3}.

Finally, FAMA computes the elemental abundances of Li, C, Na, Mg, Al, Si, Ca, Sc, Ti ({\sc i} and {\sc ii}), V, Cr, Fe ({\sc i} and {\sc ii}), Co, Ni, Y, Zr ({\sc i} and {\sc ii}), La, Ce, and Eu.

\begin{figure*}[h]
\centering
\hspace{-0.5cm}
\includegraphics[scale=0.5]{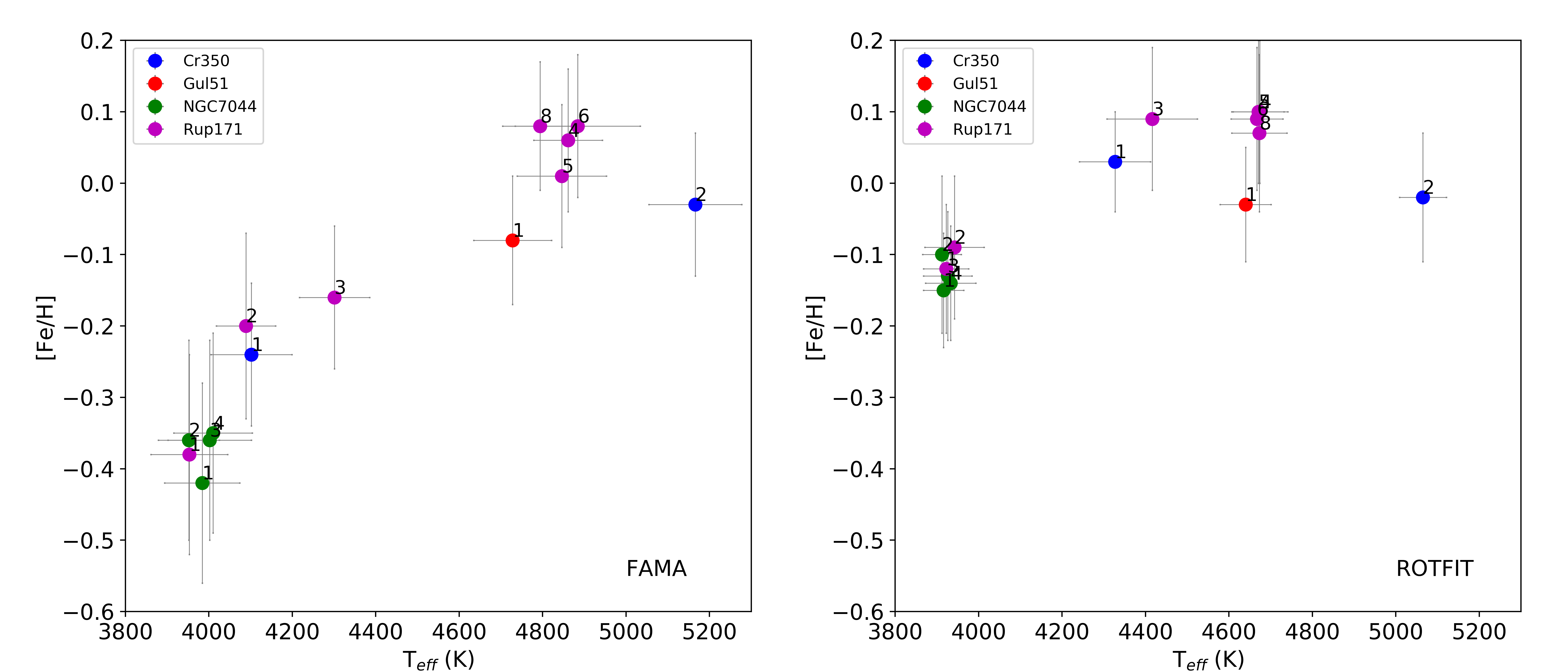}
\caption{[Fe/H] vs. $T_{\rm eff}$ for the results obtained with FAMA (left panel) and ROTFIT  (right panel). The labels indicate the ID of each star member. \label{fig:famarotfit}}
\end{figure*}

In Table~\ref{tab:clusters3} we present the results of our spectral analysis in which the four stellar parameters ($T_{\rm eff}$, $\log~g$, [Fe/H], and $\xi$) are varied up to convergence. The errors in parenthesis for [Fe/H] are the errors on the abundances generated by the uncertainties in the determination of the atmospheric parameters. One of the stars of Gulliver~51 (Gul51\_2) is a fast rotator, and therefore it cannot be analysed with FAMA;   we compute its $v \sin i$  using
ROTFIT and list this in Table~\ref{tab:ROTFIT}.
The results are presented in the left panel of Fig.~\ref{fig:famarotfit}.

\begin{table*}[ht]
\caption{Stellar parameters obtained with FAMA.}
\begin{center}
%\tiny{
\begin{tabular}{lcccc}
\hline
\hline
 ID        & $T_{\rm eff}$ & $\log~g$ & [Fe/H] & $\xi$ \\ 
          & (K) &   (dex)   & (dex)  &    (km/s)    \\
\hline
 Cr350\_1   &  4100$\pm$100    &  1.35$\pm$0.23    & $-$0.24$\pm$0.02($\pm$0.10)  &   1.76$\pm$0.10 \\
 Cr350\_2   &  5170$\pm$110   &  2.85$\pm$0.27    & $-$0.03$\pm$0.06($\pm$0.10)  &   1.58$\pm$0.09 \\
 Gul51\_1    &  4730$\pm$90    &  2.45$\pm$0.26    & $-$0.08$\pm$0.04($\pm$0.09)  &   1.52$\pm$0.08 \\
 NGC7044\_1 &  3980$\pm$90    &  0.95$\pm$0.20    & $-$0.42$\pm$0.02($\pm$0.14)  &   1.93$\pm$0.09 \\
 NGC7044\_2 &  3950$\pm$70    &  1.06$\pm$0.20    & $-$0.36$\pm$0.02($\pm$0.14)  &   1.91$\pm$0.11 \\
 NGC7044\_3 &  4000$\pm$100   &  1.23$\pm$0.21    & $-$0.36$\pm$0.03($\pm$0.14)  &   1.73$\pm$0.13 \\
 NGC7044\_4 &  4010$\pm$90    &  1.23$\pm$0.20    & $-$0.35$\pm$0.02($\pm$0.14)  &   1.82$\pm$0.10 \\
 Rup171\_1  &  3950$\pm$90    &  1.18$\pm$0.21    & $-$0.38$\pm$0.04($\pm$0.14)  &   1.66$\pm$0.12 \\
 Rup171\_2  &  4100$\pm$70    &  1.54$\pm$0.20    & $-$0.20$\pm$0.03($\pm$0.13)  &   1.57$\pm$0.08 \\
 Rup171\_3  &  4300$\pm$80    &  1.76$\pm$0.22    & $-$0.16$\pm$0.04($\pm$0.10)  &   1.59$\pm$0.10 \\
 Rup171\_4  &  4860$\pm$80    &  2.82$\pm$0.19    &    0.06$\pm$0.04($\pm$0.10)  &   1.37$\pm$0.09 \\
 Rup171\_5  &  4850$\pm$110   &  2.71$\pm$0.28    &    0.01$\pm$0.05($\pm$0.10)  &   1.46$\pm$0.08 \\
 %Rup171\_6  &  4854$\pm$111   &  2.78$\pm$0.28    &    0.01$\pm$0.05($\pm$0.08)  &   1.42$\pm$0.10 \\
 Rup171\_6  &  4880$\pm$150  &  2.91$\pm$0.20    &    0.08$\pm$0.01($\pm$0.10)  &   1.34$\pm$0.14 \\
 Rup171\_8  &  4800$\pm$90    &  2.79$\pm$0.24    &    0.08$\pm$0.05($\pm$0.09)  &   1.34$\pm$0.08 \\
 \hline
\end{tabular}
%}
%\tablefoot{}
\end{center}
\label{tab:clusters3}
\end{table*}
%\label{tab:clusters3}
%\end{table*}

From the left panel of Fig.~\ref{fig:famarotfit}, we notice a trend between metallicity and $T_{\rm eff}$. 
In particular, the coolest (lowest $\log~g$) stars of our sample reach the lowest metallicities.
For clusters with member stars spanning wide ranges in $T_{\rm eff}$ and $\log~g$
(Ruprecht~171 and Collinder~350), the trends are particularly evident: three stars of Ruprecht~171 (Rup171\_1, Rup171\_2, Rup171\_3) and one of Collinder~350 (Cr350\_1) with T$_{\rm eff}<4300$ K and $\log~g$ between 1 and 1.8 dex have much lower metallicity than the other members of the same clusters.
For clusters in which only cool giants are observed, such as NGC~7044, their [Fe/H] is lower than the literature value $-0.16$ dex \citep{Warren09}. 
Moreover, given the location of the clusters close to the Sun and their age younger than 3 Gyr, we expect a metal content near to the solar one \citep[within $\pm$0.10, considering a slope of $-$0.07~dex kpc$^{-1}$ for the radial metallicity gradient, see, e.g. ][and references therein]{zhong20}. 
This is indeed true for the hottest stars of our sample. 
For instance, the metallicity of Cr350\_2 is [Fe/H]=$-0.03\pm0.06$ in agreement within the errors with the literature value \citep[+0.03 dex,][]{blanco18}, and the mean metallicity of Rup171\_4, Rup171\_5, Rup171\_6, Rup171\_8, all RC stars, is slightly super-solar (mean value [Fe/H]$=+0.06 \pm 0.03$ dex). 

The trends with $T_{\rm eff}$ and $\log~g$ are thus general considering our sample clusters as a whole: stars located in the upper RGB, the coolest ones, are more metal-poor than stars located in RC. This topic is not new, since it has already been addressed in the analysis of cool stars on the upper RGB \citep[see, e.g.][]{worley10a,worley10b} and is discussed in more detail in Sect.~\ref{discussion}.

\subsection{Spectral analysis with ROTFIT}

We also analyse our sample stars with the code ROTFIT \citep{frasca06,frasca19}.
ROTFIT uses a grid of template spectra and performs a $\chi^{2}$ minimisation of the difference between the template and target spectrum in selected spectral regions. The grid of templates is composed of high-resolution spectra of stars with known parameters present in the ELODIE archive\footnote{\url{http://atlas.obs-hp.fr/elodie/}}. In order to use this grid, we need to degrade our HARPS-N spectra to the resolution of ELODIE (R = 42~000) and to resample them on the ELODIE spectral points ($\Delta \lambda = 0.05 \AA$).

The templates are aligned in wavelength with the analysed target spectrum through a cross-correlation function. They are broadened by convolution with a rotational profile of increasing $v \sin i$ to minimise the $\chi^{2}$. For each analysed spectral region, the weighted average of the parameters obtained for the ten best templates is taken. In particular, we analyse 28 spectral chunks, each of 100\,$\AA$ in width, in the wavelength range 4000--6800\,$\AA$. 
The final parameters ($T_{\rm eff}$, $\log~g$ and the iron abundance [Fe/H]) are the average of the results of the individual spectral region, weighted according to the $\chi^{2}$.

The atmospheric parameters, spectral type, and $v \sin i$ obtained with ROTFIT are presented in Table~\ref{tab:ROTFIT}. 
With ROTFIT, we can also provide the atmospheric parameters of Gul51\_2, the fast rotator that cannot be analysed with FAMA. It is a hot star and its $T_{\rm eff}$ and $\log~g$, derived by ROTFIT, are in agreement with the values of the  TESS catalogue \citep{tess}: $\log~g=4.12 \pm 1.54$ dex and $T_{\rm eff}=7621$ K. The metallicity of this star is
in agreement, within the errors, with the [Fe/H] of the other member star, Gul51\_1. However,  a very large error is associated with this value because of the lower $S/N$ and its fast-rotator nature, and therefore we do not consider it in the computation of the mean value of metallicity of  Gul~51.

The right panel of Fig.~\ref{fig:famarotfit} shows the metallicity [Fe/H] as a function of $T_{\rm eff}$ for the ROTFIT results. The analysis with ROTFIT yields a smaller difference in [Fe/H] between cool and warm stars of the same cluster. 
However, there are some residual differences between the cooler stars (ID 1, 2) and the hotter ones in Ruprecht~171. 
Finally, as shown by Table~\ref{tab:ROTFIT} and Fig.~\ref{fig:famarotfit}, %mettere in bold face
the metallicity obtained for NGC~7044 is in agreement with the CaT value.

\begin{table*}[ht]
\caption{Stellar parameters derived with ROTFIT.}
\begin{center}
%\tiny{
\begin{tabular}{lccccccc}
\hline
\hline
   ID         & $T_{\rm eff}$  &  $\log~g$        & [Fe/H]            &  Sp.Type   & $v \sin i$  \\%  &     RV    \\
            & (K)           &  (dex)        & (dex)             &            & (km s$^{-1}$) \\% &  (km s$^{-1}$)  \\
 \hline
 Cr350\_1     &  4330$\pm$90    & 1.28$\pm$0.24 & $ 0.03 \pm 0.07$      &     K3II   & 1.3$\pm$1.6 \\ % $ -14.57$ 
 Cr350\_2     &  5070$\pm$60    & 2.99$\pm$0.19 & $-0.02 \pm 0.09$      &    G8III   & 6.7$\pm$1.0 \\ % $ -14.73$ 
 Gul51\_1     &  4640$\pm$60    & 2.66$\pm$0.12 & $-0.03 \pm 0.08$      &    K0III   & 2.3$\pm$1.3 \\ % $ -57.36$ 
 Gul51\_2     & 7520$\pm$ 298   & 4.06$\pm$0.23 & $-0.16 \pm 0.13$      &    A9IV    & 251.6$\pm$27.2 \\  %-24.05  8.79 
 %Gul51\_2     & --     & -- & --       &    A9IV    & 251.6$\pm$27.2 \\
 NGC7044\_1   &  3920$\pm$50    & 1.53$\pm$0.16 & $-0.15 \pm 0.08$      &    K5III   & 2.0$\pm$1.8 \\ % $ -51.36$ 
 NGC7044\_2   &  3910$\pm$50    & 1.51$\pm$0.16 & $-0.10 \pm 0.11$      &    K5III   & 1.8$\pm$1.8 \\ % $ -49.90$ 
 NGC7044\_3   &  3930$\pm$60    & 1.50$\pm$0.19 & $-0.13 \pm 0.09$      &    K5III   & 1.7$\pm$1.7 \\ % $ -47.75$ 
 NGC7044\_4   &  3930$\pm$60    & 1.50$\pm$0.17 & $-0.14 \pm 0.08$      &    K5III   & 2.0$\pm$1.8 \\ % $ -49.17$ 
 Rup171\_1  &  3920$\pm$50      & 1.56$\pm$0.13 & $-0.12 \pm 0.09$      &    K5III   & 2.0$\pm$1.6 \\ % $ 213.20$ 
 Rup171\_2  &  3940$\pm$70      & 1.56$\pm$0.14 & $-0.09 \pm 0.10$      & K3.5IIIb   & 1.6$\pm$1.6 \\ % $   5.23$ 
 Rup171\_3  &  4420$\pm$110     & 2.13$\pm$0.43 & $ 0.09 \pm 0.10$      &    K2III   & 1.4$\pm$1.5 \\ % $ -93.29$ 
 Rup171\_4  &  4670$\pm$70      & 2.67$\pm$0.12 & $ 0.10 \pm 0.10$      &  K1.5III   & 0.7$\pm$0.9 \\ % $   6.39$ 
 Rup171\_5  &  4670$\pm$60      & 2.66$\pm$0.13 & $ 0.10 \pm 0.10$      &  K1.5III   & 0.8$\pm$1.1 \\ % $   0.00$ 
 Rup171\_6  &  4670$\pm$60      & 2.64$\pm$0.14 & $ 0.09 \pm 0.10$      &    K0III   & 1.0$\pm$1.2 \\ % $   0.00$ 
 Rup171\_8  &  4670$\pm$60      & 2.64$\pm$0.14 & $0.07  \pm0.11$       &  K1.5III   & 0.7$\pm$0.9 \\ % $   0.00$ 
 \hline
\end{tabular}
%}
%\tablefoot{}
\end{center}
\label{tab:ROTFIT}
\end{table*}

\section{The cool giant stars}
\label{discussion}
In this section,  we investigate the causes of the low metallicities measured with the EWs in the cool giants.
In principle, stars in open clusters should present a homogeneous chemical composition \citep[see, e.g.][]{bovy16}, at least within some range, typically of few $\sim$0.01~dex \citep[cf.][for examples of the possible presence of some degree of inhomogeneity in open clusters]{liu16, spinaplei18}. However, stars belonging to the same open cluster, but in different evolutionary stages, might display more notable differences in their chemical patterns. These differences can be due to physical phenomena, such as atomic diffusion and mixing \citep[e.g.][]{lagarde18, casali19,  bertellimotta17, semenova},
or to analysis effects, such as non-local thermodynamic equilibrium (NLTE) effects or correlations between atmospheric parameters and abundances \citep[see, e.g.][for a review]{blanco15}. 
In addition, the analysis of cool giant stars can be affected by several complications, such as the presence of a forest of molecular lines, possible asymmetric shapes of the lines due to mass loss, deviations from hydrostatic equilibrium, the presence of giant convective cells, and the deviations from LTE \citep[see, e.g.][]{asplund05, bergemann14}. 
Here we discuss some aspects of the cool giant analysis, including the choice of the stellar parameters, the correction for NLTE, the selection of the line list, the adoption of the atmosphere models, and the continuum placement.

\subsection{The role of stellar parameters}
We investigate the effect of adopting stellar parameters from {\em Gaia}, which are calculated based on their photometry, rather than the spectroscopic ones with FAMA which are based on the EW analysis.
If there is indeed a deviation from the hydrostatic equilibrium, the gravity derived from the ionisation balance might be incorrect, and thus produce incorrect abundances.
The simultaneous determination of the stellar parameters from spectroscopy can produce, for instance,  a degeneration among those parameters, producing a correlation between $\log~g$ and [Fe/H] \citep[cf.][]{bc14}. 
Therefore, the use of temperature and $\log~g$ independent of spectroscopy might solve this eventual degeneracy. 
%We do not expect large variation in the metallicities computed with the photometric parameters, since the photometric stellar parameters are, on average, in good agreement with spectroscopic ones (see Tables~\ref{tab:input} and \ref{tab:clusters3}). 
First of all, we fix the gravity to the values of Table~\ref{tab:input} using both the gravities from {\it Gaia} and from the comparison with isochrones, and let $T_{\rm eff}$, $\xi$, and [Fe/H] vary up to convergence. We then  keep the photometric $\log~g$ and $T_{\rm eff}$ constant, while $\xi$ and [Fe/H] are varied up to convergence. 
Even with these choices, the global trends of [Fe/H] versus $T_{\rm eff}$, with cool and low-gravity stars having a lower metal content than the other member stars, are still present with the EW method (e.g. a discrepancy in metallicity of $\sim0.3$ and $\sim0.2$ dex, respectively, between the coolest and hottest star of Rup~171).

\subsection{The NLTE effects}
FAMA uses MOOG to compute abundances in the LTE approximation. The photospheres of cool giants, with their low surface gravities and thus low densities, might depart from LTE, being translucent over large radial extensions. Thus, the radiative rates can dominate the collisional rates for many atomic transitions \citep[cf.][]{short03}. This effect is usually in place in low-gravity giant stars, but it is stronger for metal-poor stars. 
The departure from LTE could be due to the high excitation levels of Fe\,{\sc i}, which do not thermally couple to the ground state of Fe\,{\sc ii}.
Another aspect of the departure could be the treatment of poorly known inelastic collisions with hydrogen atoms \citep{mashonkina11}. 
There are indeed several studies showing that NLTE effects in the ionisation balance of Fe\,{\sc i}/Fe\,{\sc ii} are larger for giant  metal-poor stars \citep[e.g.][and references therein]{bergemann12,mashonkina11,collet05} than for their counterparts at higher metallicity.
Since iron lines are used to derive stellar parameters, we aim to estimate the effect of NLTE in the spectra of our giant stars, even if their metallicity is not so low that we expect a strong departure from LTE.
%, and the consequences on the determination of the  parameters. 
%This causes a negligible differences between the NLTE and LTE abundances inferred from the Fe~I lines.
%Iron plays a relevant role in studies of cool stars because of its many lines in the visible spectrum. 
%Therefore, a NLTE effect can affect the determination of the atmospheric parameters and then the abundances in cool or metal-poor stars.
%In our case, we want to test the NLTE effects for cool giant stars in solar metallicity range.

We estimate the NLTE abundance corrections for each Fe line using the calculator by the MPIA NLTE group\footnote{\url{http://nlte.mpia.de/gui-siuAC_secE.php}} \citep{bergemann12}. The NLTE abundance corrections are computed as $\Delta \rm Fe = \log A(Fe)_{NLTE} - \log A(Fe)_{LTE}$, which is the difference between the NLTE and LTE abundances.
These corrections on the considered Fe lines are of the order $0.001-0.02$ dex. They are negligible as expected for solar metallicity giants \citep[][]{bergemann12}, even at the very low surface gravities of the coolest stars of our sample.
Thus, these corrections alone cannot justify the discrepancy in metallicity among members of the same cluster, and they do not affect the determination of the spectroscopic stellar parameters from the EW analysis.

\subsection{The line list}

The spectra of cool stars, with $T_{\rm eff} <$ 5000~K, are characterised by strong line crowding and consequent blending. However, HARPS-N has high spectral resolution, R=115~000, which makes the problem of the blending  less dominant.  As pointed out by \citet{tsantaki13}, the identification of unblended EWs of photospheric lines and the continuum placement are  difficult tasks in cool stars. This might lead to incorrect measurement of the EWs with consequent errors in the derived stellar parameters. Following the work of \citet{tsantaki13}, we adopt their line list designed for the analysis of cool stars. However, we reiterate that the line list of \citet{tsantaki13} is designed for cool solar-type stars. In cool giant stars, the effect of crowding and continuum placement can be even more severe. 

To understand whether or not the difference in metallicity is due to the Gaia-ESO line list (used to obtain the results in Table~\ref{tab:clusters3}), we perform the EW spectral analysis using the line list by \citet{tsantaki13}. The atmospheric parameters achieved by both line lists are consistent within the errors. Moreover, there are still differences between the [Fe/H] in stars in distinct evolutionary stages within the same cluster.

\subsection{The model atmospheres}
Finally, we test the influence of the choice of model atmospheres on the determination of stellar parameters and abundances. To perform our test, we recompute the stellar parameters using the Kurucz models \citep{castelli03} instead of the MARCS spherical models. The differences are negligible for all parameters.  
However, as discussed by \citet{short03}, for giants of spectral types G to M, the failure in reproducing the relation between $T_{\rm eff}$ and the colours in the blue and violet regions of the spectrum, such as $B-V$, is typical of most model atmospheres. \citet{bessel98} indicated the incomplete or erroneous opacity in the blue-violet region as the origin of that discrepancy. 
Since this problem affects different types of model atmospheres, the change of model does not help to solve the discrepancy. %and it might be one of the causes at the origin of the perennial problem of the metallicity determination in cool giants. 

\subsection{Suggestions on [Fe/H] determination in cool stars}

The problem of obtaining overly low metallicity in cool stars has been known for a long time, but  a clear solution is still missing. %\citep{worley10a,worley10b}. 
An example of this effect in open clusters is the star at the tip of the RGB of the cluster Collinder~261. This star is cooler than the other studied members, and its measured metal content is lower than the other stars \citep{carretta05,friel03}.   

There is not a single explanation to clarify the differences in metallicities in stars of the same cluster belonging to different evolutionary phases, in particular the upper RGB. For the cluster with a larger number of observed member stars, Rup~171,
we severely underestimate [Fe/H] with the EW method (maximum differences $>0.4$~dex) in stars cooler than $\sim 4300$~K and with $\log~g<1.8$ dex.
This effect is smaller with parameters and metallicities derived through ROTFIT. In this case, the maximum differences in [Fe/H] between the coolest and hottest stars are of the order of $\sim$0.2~dex. 
%{\bf In addition, the effect is mitigated by the use of photometric surface gravities, indicating that the classical method used to derive $\log~g$ from spectroscopy, i.e. the ionization balance, is failing for these stars.} 
For NGC~7044, we are limited by the observations of four member stars, all around the RGB tip, and therefore we cannot compare the cool stars with the warm ones.  From them, it is difficult to evaluate whether or not their [Fe/H] values are underestimated. However, the [Fe/H] measured with ROTFIT is higher by about 0.2~dex with respect to [Fe/H] determined with the EWs. In addition, the former metallicity is in better agreement with the literature value (albeit based on the CaT method, which is not free from large uncertainties and biases). 
 We analyse two member stars of Cr~350. For the hottest one, spectral fitting and EW methods are in agreement, while for the coolest one the EW method tends to underestimate its [Fe/H], as in the other cool giants. 

%{\bf Fixing the surface gravity to its photometric value improves the result for the coolest star.} Finally, in Gul~51 we analyse a single star, with a relatively high $\log~g$, in which both methods are in agreement for its [Fe/H]. 

\begin{figure}[h]
\centering
\includegraphics[scale=0.5]{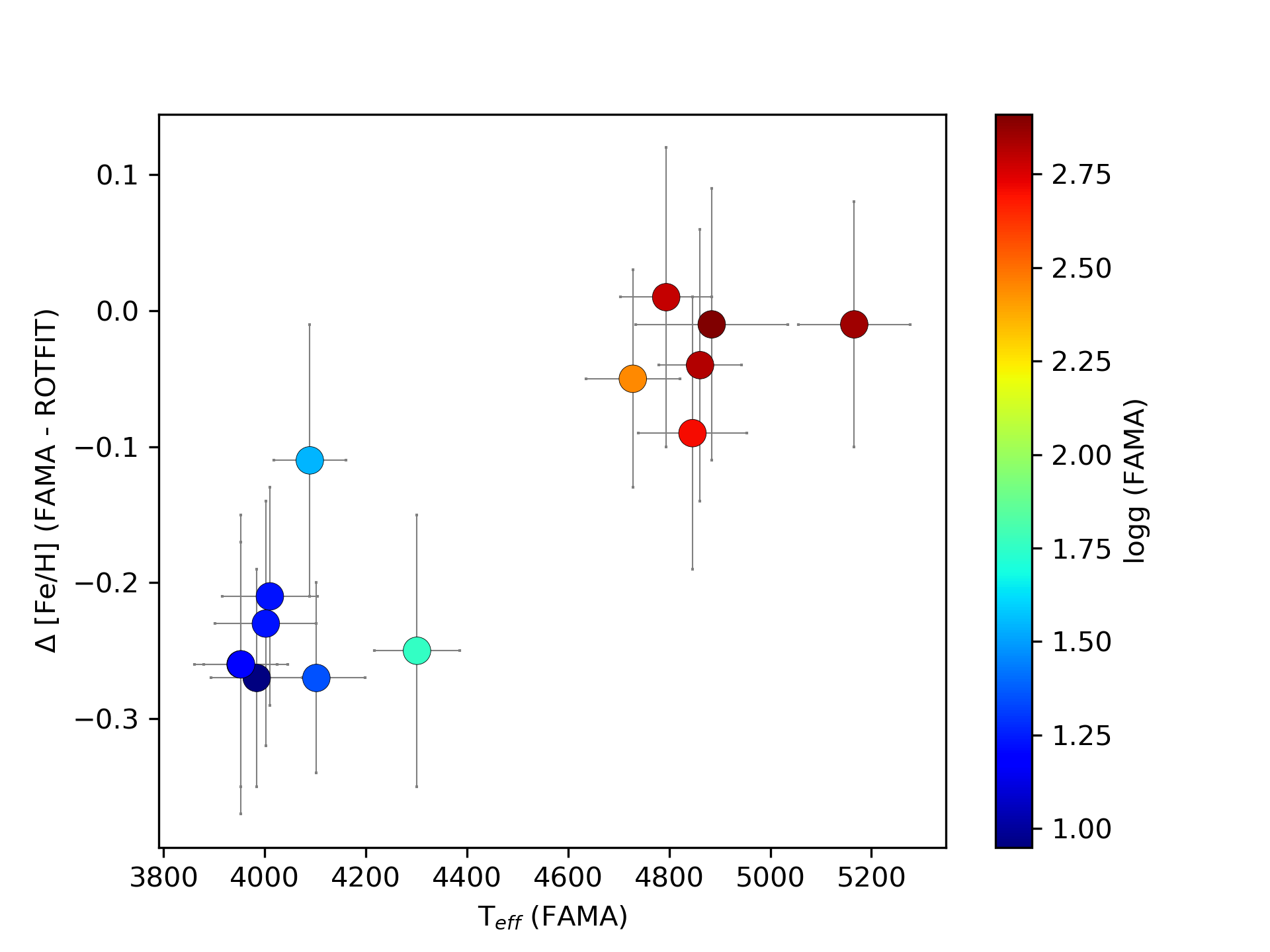} 
\caption{Difference in [Fe/H] between FAMA and ROTFIT as a function of $T_{\rm eff}$, colour-coded by $\log~g$. \label{fig:deltafeh} }
\end{figure}

\begin{figure*}[h]
\centering
\includegraphics[scale=0.5]{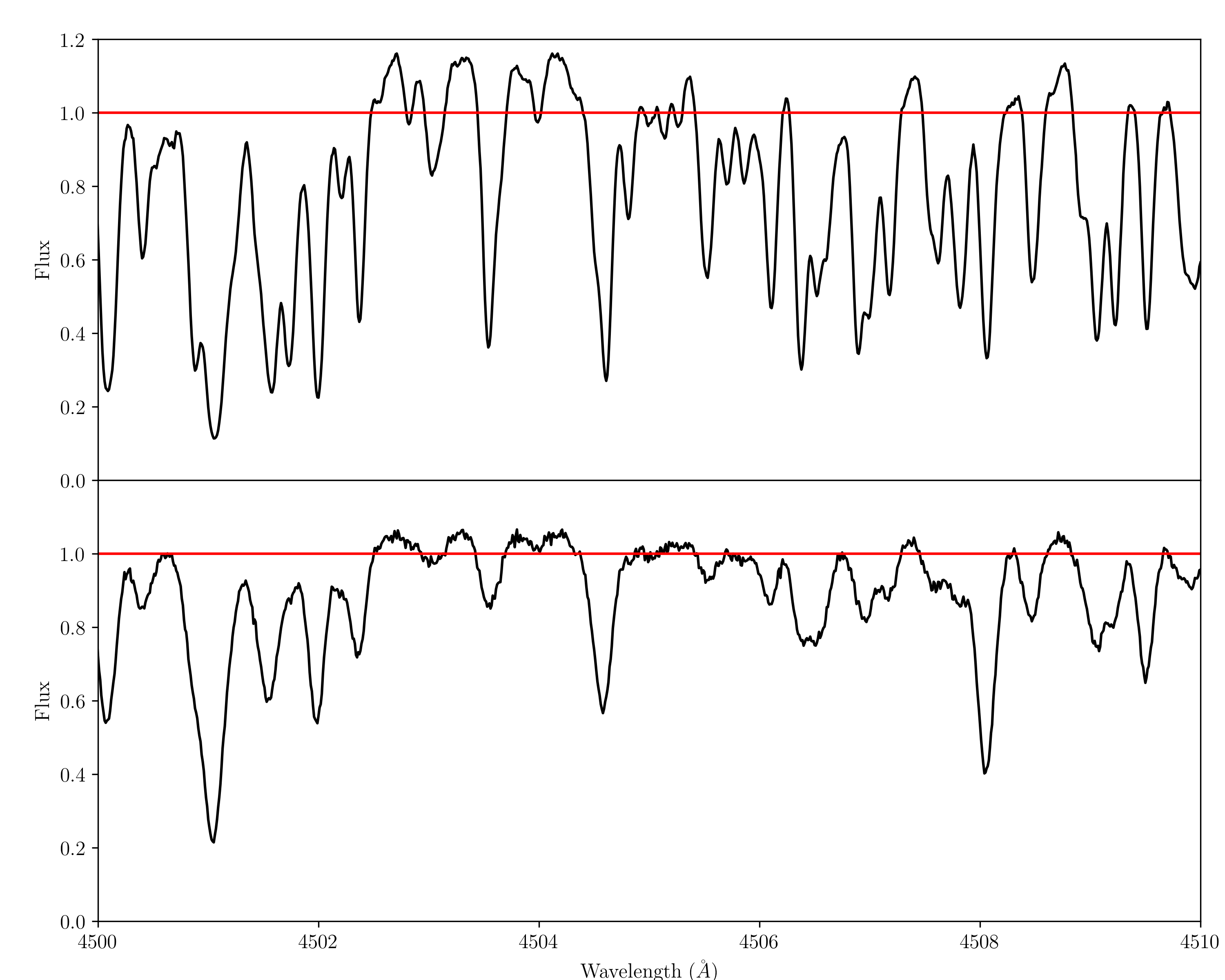} 
\caption{Portion of the spectra, normalised with {\sc Daospec}, of the cool star Cr350\_1 (top panel) and the warm star Cr305\_2 (bottom panel) with the continuum (red line). \label{fig:spectrum} }
\end{figure*}

The combination of several aspects makes the determination of metallicity from the EW analysis in the cool giants (T$_{\rm eff} < 4300$ K, $\log~g<1.8$ dex) unreliable, as shown in Fig.~\ref{fig:deltafeh}, where we present the difference in [Fe/H] between FAMA and ROTFIT as a function of $T_{\rm eff}$, colour-coded by $\log~g$. 
From our analysis described in the previous sections, we consider the principal causes to be as follows:  the erroneous opacity in  model atmospheres of cool G to M giant stars \citep[see, e.g][]{bessel98, short03}, and the large number of lines in the spectrum, both atomic ones and molecular bands for cool and metal rich stars,
which makes it difficult to define the continuum near to the lines of interest and thus to measure reliable EWs. Regarding the continuum placement, we recall that {\sc Daospec} adopts a global and not local continuum and this aspect can only increase the difficulty in the continuum setting for cool giants.
A clear example of this problem is seen in Fig.~\ref{fig:spectrum}, where two normalised spectra are compared: the spectrum of the cool star Cr350\_1 (top panel) and the spectrum of the warm star Cr350\_2 (bottom panel). The continuum computed by {\sc Daospec} for Cr350\_1 produces a normalised spectrum slightly above 1, in which the EWs are underestimated, in contrast to Cr350\_2. 
As explained in \citet{cantat14}, the continuum defined by {\sc Daospec}, on which the EW fit is based, is not the true continuum of the spectrum (i.e. the continuous star emission after all the lines are excluded), but an effective continuum, which is the true continuum depressed by a statistical estimate of the contaminating lines (the unresolved or undetected ones, producing a sort of line blanketing). The use of the effective continuum improves the measurement of unblended lines, as demonstrated in \citet{cantat14}, but it is sometimes perceived as being too low, especially in spectra dominated by line crowding (i.e. in particular high-metallicity giant stars) or with decreasing $S/N$ of the spectra.
The case of  Cr350\_1 is a typical example of our limits in measuring the EWs of metal-rich giant stars, for which we likely underestimate the EWs and, consequently, we derive a lower [Fe/H] than in warmer stars.
Moreover, it is known also from other works that the analysis based on EW measurements tends to underestimate the [Fe/H] of cool giant stars, as shown by the EW analyses of benchmark stars performed by different groups \citep{jofre14,heiter2015}. %However, we recall that there are no benchmark stars with such low temperatures and gravities, and thus such regime remains quite unexplored. 
Consequently, for the spectra of stars cooler than 4300~K, ROTFIT, which is less prone to continuum setting and blending effects, produces more solid determinations of the stellar parameters than the EW analysis. 

%In addition, for most of our cool giants the S/N is quite low, making even more difficult the definition of the local continuum.

In what follows, we consider the results from the EW analysis only for stars hotter than 4300 K, namely one star in Cr~350 and in Gul~51, and four stars in Rup~171. 
In Table~\ref{tab:ave:met}, we compare the mean metallicities of our cluster sample from ROTFIT (all sample) and from FAMA (only stars with $T_{\rm eff}> 4300$ K). The number of stars used to compute the average [Fe/H] is reported in parentheses. 
The two determinations of the mean cluster metallicity are in good agreement within the uncertainties (1-$\sigma$ standard deviation). 

\begin{table}[ht]
\caption{Mean metallicities of our sample open clusters.}
\begin{center}
\fontsize{9pt}{9pt}
\selectfont
%\tiny{
\begin{tabular}{lcc}
\hline
\hline
Cluster       &   [Fe/H] ROTFIT (all)        & [Fe/H] FAMA (warm)   \\ 
               
\hline
Collinder~350  &     0.00$\pm$0.08 (2) & $-$0.03$\pm$0.08 (1)\\
Gulliver~51    &  $-$0.03$\pm$0.08 (1) & $-$0.08$\pm$0.04 (1)\\
NGC~7044       &  $-$0.13$\pm$0.02 (4) & --\\                                 
%Ruprecht~171   &     0.03$\pm$0.09 (7) & 0.04$\pm$0.03 (4)\\
Ruprecht~171   &     0.03$\pm$0.09 (7) & 0.06$\pm$0.03 (4)\\
 \hline \hline
\end{tabular}
%}
\end{center}
\label{tab:ave:met}
\end{table}

\section{Chemical abundances}
\label{abu}
Elemental abundances are computed with FAMA using the routines {\sc abfind} and {\sc blends} of MOOG. The latter is used for elements 
that present hyperfine splitting in their lines. 
To compute the Solar-scaled abundances and abundance ratios [X/H] and [X/Fe], we define our Solar scale measuring the element abundances on a solar spectrum. For this task we use a spectrum of Ceres collected by the twin HARPS spectrograph at the 3.6 m ESO telescope. The Solar abundances are listed in Table~\ref{tab:sun}, in which we show our Solar abundances and the photospheric Solar abundances from \citet{grevesse07}. %and the Solar abundances for neutron capture elements derived by \citet{magrini18} in giant stars of M67. 
For the elements that cannot be measured in the Ceres spectrum, we use the values from \citet{grevesse07}. %For La, we adopt the value from \citet{magrini18} computed using the giants of M67.

\begin{table}[ht]
\caption{Solar chemical abundances.}
\begin{center}
\begin{tabular}{lcc}%c}
\hline
\hline
Element & Sun (Ceres) & Sun (G07) \\% &  M~67 giants (M18)\\
\hline
Li\,{\sc i}        &       --           &  $1.05 \pm 0.10$ \\ % & --\\          
C\,{\sc i}         &       --           &  $8.39 \pm 0.05$ \\ % & -- \\  %CERES:$8.16\pm0.06$
Na\,{\sc i}        &    $6.22\pm0.04$   &  $6.17 \pm 0.04$ \\ % & -- \\
Mg\,{\sc i}        &    $7.63\pm0.02$   &  $7.53 \pm 0.09$ \\ % & -- \\
Al\,{\sc i}        &    $6.41\pm0.02$   &  $6.37 \pm 0.06$ \\ % & --  \\
Si\,{\sc i}        &    $7.45\pm0.04$   &  $7.51 \pm 0.04$ \\ % & -- \\
Ca\,{\sc i}        &    $6.29\pm0.05$   &  $6.31 \pm 0.04$ \\ % & -- \\
Sc\,{\sc ii}       &    $3.17\pm0.03$   &  $3.17 \pm 0.10$ \\ % & -- \\
Ti\,{\sc i}        &    $4.87\pm0.05$   &  $4.90 \pm 0.06$ \\ % & -- \\
Ti\,{\sc ii}       &    $4.99\pm0.07$   &     --           \\ % & -- \\
V\,{\sc i}         &    $3.92\pm0.07$   &  $4.00 \pm 0.02$ \\ % & -- \\
Cr\,{\sc i}        &    $5.62\pm0.03$   &  $5.64 \pm 0.10$ \\ % & -- \\
Fe\,{\sc i}        &    $7.45\pm0.09$   &  $7.45 \pm 0.05$ \\ % & --  \\
Fe\,{\sc ii}       &    $7.47\pm0.09$   &     --           \\ % & -- \\
Co\,{\sc i}        &    $4.89\pm0.08$   &  $4.92 \pm 0.08$ \\ % & -- \\
Ni\,{\sc i}        &    $6.23\pm0.08$   &  $6.23 \pm 0.04$ \\ % & -- \\
Y\,{\sc ii}        &    $2.11\pm0.09$   &  $2.21 \pm 0.02$ \\ % &  $2.14 \pm 0.01$ \\
Zr\,{\sc i}        &      --            &  $2.58 \pm 0.02$ \\ % &  -- \\
Zr\,{\sc ii}       &    $2.42\pm0.06$   &     --           \\ % &  $2.54 \pm 0.03$ \\
La\,{\sc ii}       &    $1.13\pm0.36$   &  $1.13 \pm0.05 $ \\ % &  $1.00 \pm 0.02$  \\ %$1.13 \pm 0.05$
Ce\,{\sc ii}       &     --             &  $1.70 \pm 0.10$ \\ % &  $1.71 \pm 0.01$ \\
Eu\,{\sc ii}       &     --             &  $0.52 \pm 0.06$ \\ % &  $0.42 \pm 0.01$  \\
\hline
\end{tabular}
\tablefoot{G07: \citet{grevesse07}.} %, M18: \citet{magrini18}}
\end{center}
\label{tab:sun}
\end{table}

\begin{table*}[ht]
\caption{Elemental abundances of each star with T$_{\rm eff}> 4300$ K, in the form 12 + log(X/H).}
\begin{center}
\small{
\begin{tabular}{lccccccccccc}
\hline
\hline
  Star    & A(Li\,\textsc{i}) & n(Li\,{\sc i}) & A(C\,{\sc i}) &  n(C\,{\sc i})  & A(Na\,{\sc i}) &  n(Na\,{\sc i})  &     A(Mg\,{\sc i}) &  n(Mg\,{\sc i})  &    A(Al\,{\sc i}) &  n(Al\,{\sc i})  & ...\\
\hline
Cr350\_2   & $1.41 \pm 0.03 $ &  1  & $7.88 \pm 0.07$ & 2 & $6.43 \pm 0.02$  & 2 & $7.64 \pm 0.12   $ & 2 &  $6.36 \pm 0.01  $ & 2 & \\
Gul51\_1   & $1.05 \pm 0.07 $ &  2  & $7.92 \pm 0.06$ & 1 & $6.26 \pm 0.07$  & 4 & $7.62 \pm 0.01   $ & 2 &  $6.33 \pm 0.03  $ & 2 & \\
Rup171\_4  & --               &  -- & $7.98 \pm 0.03$ & 2 & $6.41 \pm 0.07$  & 4 & $7.73 \pm 0.02   $ & 3 &  $6.51 \pm 0.03  $ & 2 & \\
Rup171\_5  & --               &  -- & $7.81 \pm 0.08$ & 1 & $6.43 \pm 0.06$  & 4 & $7.71 \pm 0.01 $ & 2 & $6.47 \pm 0.01 $ & 2 & \\
Rup171\_6  & --               &  -- & $7.99 \pm 0.06$ & 2 & $6.42 \pm 0.03$  & 4 & $7.67 \pm 0.02   $ & 2 &  $6.45 \pm 0.07  $ & 2 & \\
Rup171\_8  & $0.39 \pm 0.06 $ &  1  & $7.92 \pm 0.02$ & 2 & $6.42 \pm 0.06$  & 4 & $7.70  \pm 0.01$ & 2 &  $6.50 \pm 0.04  $ & 1 & \\
\hline
\end{tabular}
}
\tablefoot{n(X) represents the number of lines for each element. The full version of this table is available online at the CDS. }
\end{center}
\label{tab:abustars}
\end{table*}

We cannot produce
abundances for NGC~7044 and for the cool stars of Rup~171 and Cr~350 with the stellar parameters derived via EW analysis. 
In Table~\ref{tab:abustars} we list the elemental abundances for each warm member star ($T_{\rm eff} > 4300$ K), where the errors are the standard deviations on the mean of the abundances of each line for the given elements.
When only one line is detected per element, the error is estimated propagating the uncertainty on its EW. The uncertainties reported in Table~\ref{tab:abustars} do not take into account the impact of the uncertainties on the stellar parameters, and are therefore lower limits in the total error budget. Typical errors due to stellar parameters are estimated for [Fe/H]. They are reported in Table~\ref{tab:clusters3} and range from 0.01 to 0.06~dex. 

In Table~\ref{tab:abu}, 
we report the mean abundance ratios for each cluster. 
The errors for Rup~171 are the standard deviations of the mean, as we have four warm stars for these clusters; whereas, for Gul~51 and Cr~350 we can use only one star, and therefore the uncertainties are the errors on each individual measurement.

\begin{table*}[ht]
\caption{Mean abundance ratios for the clusters.}
\begin{center}
%\fontsize{9pt}{9pt}
\small{
\begin{tabular}{lrrr}
\hline
\hline
Ratio & Cr~350 & Gul~51 &  Rup~171 \\
\hline
\hline 
odd-elements        &                  &       & \\
\hline 
$ \rm [Na/Fe]    $ &  $ 0.25 \pm 0.10$   &  $ 0.12 \pm 0.12$        &    $0.14\pm 0.03$  \\ 
$ \rm [Al/Fe]        $ &  $-0.01 \pm 0.10$   &     $0.00\pm0.09$    &   $0.02 \pm 0.03$  \\
\hline 
$\alpha$-elements        &                  &       & \\
\hline 
%$ \rm [O/Fe]         $ &  $-0.16 \pm 0.11$   &  $-0.10 \pm 0.14$    &   $-0.06 \pm 0.08$  \\
$ \rm [Mg/Fe]        $ &  $ 0.05 \pm 0.15$   &  $ 0.07 \pm 0.09$    &    $0.02 \pm 0.04$  \\
$ \rm [Si/Fe]        $ &  $ 0.03 \pm 0.12$   &  $ 0.07 \pm 0.10$    &    $0.03 \pm 0.02$  \\
$ \rm [Ca/Fe]        $ &  $-0.01 \pm 0.12$   &  $-0.02 \pm 0.10$    &   $-0.01 \pm 0.02$  \\
$ \rm [Sc/Fe]        $ &  $ 0.00 \pm 0.13$   &  $ 0.06 \pm 0.11$    &    $0.09 \pm 0.02$  \\
$ \rm [Ti/Fe]^{**}   $ &  $ 0.00 \pm 0.14$   &  $-0.01 \pm 0.13$    &    $0.02 \pm 0.03$  \\
\hline 
iron-peak elements        &                  &       & \\
\hline 
$ \rm [V/Fe]         $ &  $ 0.00 \pm 0.15$   &  $ 0.04 \pm 0.14$     &    $0.10 \pm 0.03$  \\ 
$ \rm [Cr/Fe]        $ &  $ 0.02 \pm 0.13$   &  $-0.01 \pm 0.11$     &   $-0.01 \pm 0.02$  \\
$ \rm [Co/Fe]        $ &  $-0.02 \pm 0.14$   &  $-0.02 \pm 0.14$     &    $0.10 \pm 0.02$  \\
$ \rm [Ni/Fe]        $ &  $-0.06 \pm 0.13$   &  $-0.05 \pm 0.12$     &    $0.03 \pm 0.02$  \\
\hline 
neutron-capture elements        &                  &       & \\
\hline 
$ \rm [Y/Fe]         $ &  $ 0.04 \pm 0.16$   &     $ 0.08 \pm 0.12$   &    $0.01\pm 0.03$  \\
$ \rm [Zr/Fe]^{*,**}   $ &  $ 0.13 \pm 0.14$   &     $ 0.21 \pm 0.12$   &    $0.09\pm 0.04$  \\ 
$ \rm [La/Fe]        $ &  $ -0.13 \pm 0.13$   &     $ -0.07 \pm 0.12$   & $-0.22\pm 0.03$  \\
$ \rm [Ce/Fe]^{*}    $ &  $-0.04 \pm 0.14$   &     $ 0.04 \pm 0.14$   & $-0.04\pm 0.03$  \\ 
$ \rm [Eu/Fe]^{*}    $ &  $-0.03 \pm 0.12$   &     $ 0.08 \pm 0.29$   & $-0.02\pm 0.06$  \\ 

 \hline
\end{tabular}
}
\tablefoot{Abundance ratios on our Solar scale, with the exception of ($^{*}$), which are calculated using the solar value from \citet{grevesse07}. 
($^{**}$) indicates the average value between [TiI/Fe] and [TiII/Fe], and [ZrI/Fe] and [ZrII/Fe].}
\end{center}
\label{tab:abu}
\end{table*}

\subsection{Comparison with the literature}
 
\begin{figure}[h]
\centering
\hspace{-0.5cm}
\includegraphics[scale=0.55]{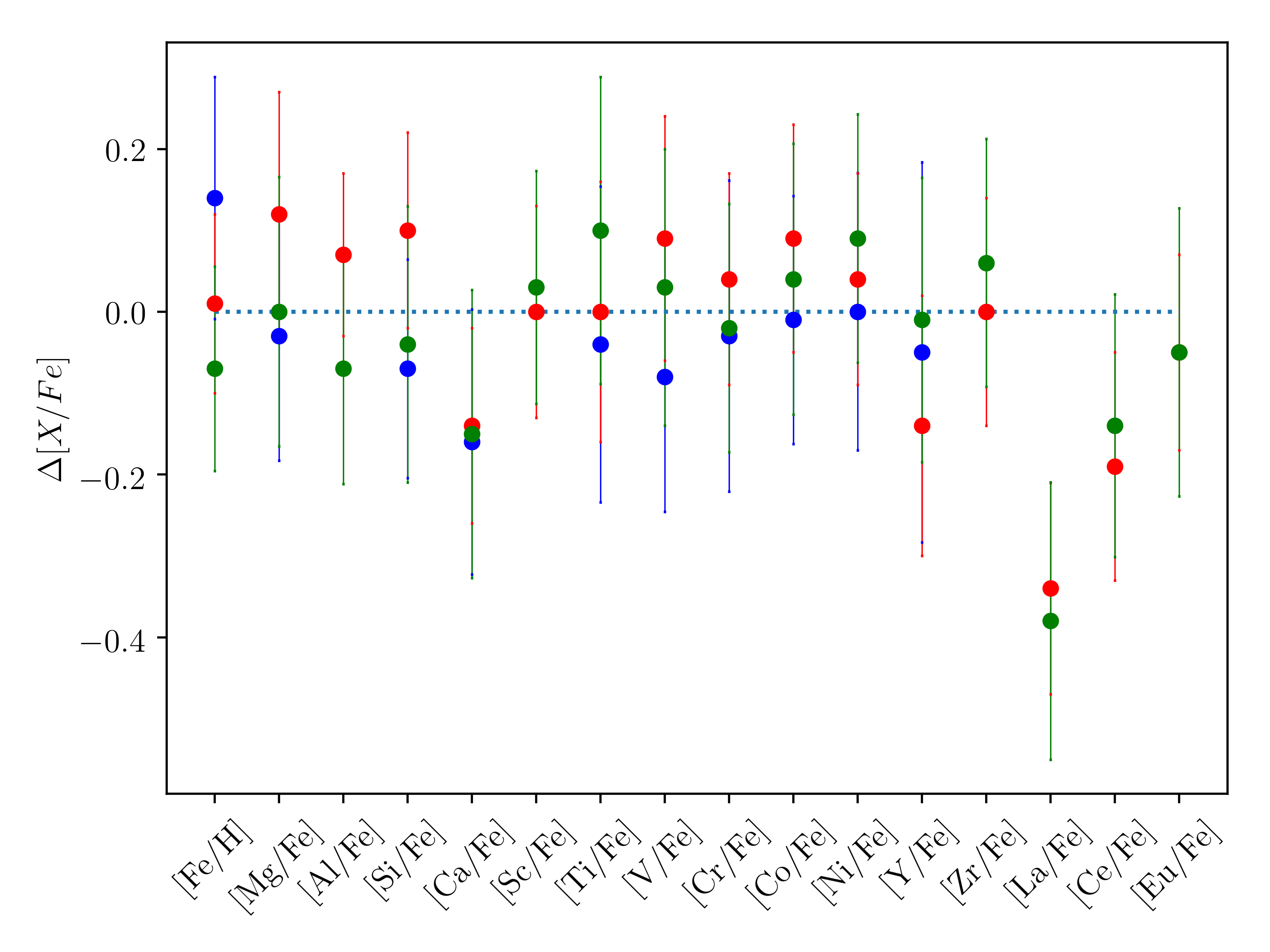} 
\caption{Difference between our abundance ratios and the values from the literature: in green the comparison with \citet{pakhomov09}, in blue the comparison with \citet{blanco15}, and in red the comparison with \citet{blanco18}. In the last case, the error bars take into account only the errors on the abundance ratios computed in this work, and therefore they are lower limits.  \label{fig:cr350_confronto} }
\end{figure}

Collinder~350 is the only cluster of our sample that has previous determinations of chemical abundances from high-resolution spectral analysis. The previous works on this cluster presenting high-resolution spectroscopy are \citet{pakhomov09}, \citet{blanco15}, and \citet{blanco18}. The first work analysed a spectrum of the star Cr350\_1 collected with the red branch spectrograph  ($R\sim 50~000$) at the 2.16 m Telescope at the Xinglong Observatory in China. \citet{pakhomov09} found a metallicity of [Fe/H]=$+0.11 \pm 0.06$ dex. \citet{blanco15} and \citet{blanco18} analysed an archival NARVAL spectrum of a giant star ($R\sim 80~000$), obtaining slightly different values of [Fe/H]=$-0.10 \pm 0.01$ (or [Fe/H]=$0.0$ with different normalisation) and 0.03 dex, respectively. 
 Table~\ref{tab:cr350} shows the abundance ratios for Cr350\_1 by \citet{pakhomov09} and for a giant star in Collinder~350, for which the coordinates are not available in the papers \citep{blanco15,blanco18}, compared with our results of Cr350\_2. We exclude from the comparison the elements that are modified by stellar mixing, such as C and Na, because we do not compare the same star.  
Also, without taking into account the possible differences
in solar reference abundances adopted in the studies mentioned above, Fig.~\ref{fig:cr350_confronto} shows that most of our abundance ratios are in agreement within the errors between our work and those papers. The only exception is [La/Fe], for which we find a lower value.

\begin{table}[ht]
\caption{Abundances for Collinder~350}
\begin{center}
\small{
\begin{tabular}{lrrrr}
\hline
\hline

 Ratio   &  BC18  &  BC15  & P09  &    This work \\ 
\hline
$ \rm [Fe/H] $  &  $0.03 $  &   $-0.10 \pm 0.01 $ &  $ 0.11 \pm 0.06$  & $0.04  \pm  0.11$  \\
$ \rm [Mg/Fe]$  &  $-0.07$  &   $0.08  \pm 0.03 $ &  $ 0.05 \pm 0.07$  & $0.05  \pm  0.15$  \\
$ \rm [Al/Fe]$  &  $-0.08$  &   --                &  $ 0.06 \pm 0.10$  & $-0.01 \pm  0.10$  \\
$ \rm [Si/Fe]$  &  $-0.07$  &   $0.10  \pm 0.06 $ &  $ 0.07 \pm 0.12$  & $0.03  \pm  0.12$  \\
$ \rm [Ca/Fe]$  &  $0.04 $  &   $0.06  \pm 0.11 $ &  $ 0.05 \pm 0.13$  & $-0.10 \pm  0.12$  \\
$ \rm [Sc/Fe]$  &  $0.00 $  &   --                &  $-0.03 \pm 0.06^{*}$  & $0.00  \pm  0.13$  \\ 
$ \rm [Ti/Fe]$  &  $-0.02$  &   $0.02  \pm 0.11 $ &  $-0.12 \pm 0.10$  & $-0.02 \pm  0.16$  \\  
$ \rm [V/Fe] $  &  $-0.09$  &   $0.08  \pm 0.07 $ &  $-0.03 \pm 0.08$  & $0.00  \pm  0.15$  \\        
$ \rm [Cr/Fe]$  &  $-0.02$  &   $0.05  \pm 0.14 $ &  $ 0.04 \pm 0.08$  & $0.02  \pm  0.13$  \\
$ \rm [Co/Fe]$  &  $-0.11$  &   $-0.01 \pm 0.06 $ &  $-0.06 \pm 0.09$  & $-0.02 \pm  0.14$  \\
$ \rm [Ni/Fe]$  &  $-0.10$  &   $-0.06 \pm 0.11 $ &  $-0.15 \pm 0.08$  & $-0.06 \pm  0.13$  \\
$ \rm [Y/Fe] $  &  $0.18 $  &   $0.09  \pm 0.17 $ &  $ 0.05 \pm 0.07$  & $0.04  \pm  0.16$  \\  
$ \rm [Zr/Fe]$  &  $0.13 $  &   --                &  $ 0.07 \pm 0.06^{*}$  & $0.13  \pm  0.14$  \\
$ \rm [La/Fe]$  &  $0.21 $  &   --                &  $ 0.25 \pm 0.11$  & $-0.13  \pm  0.13$  \\                      
$ \rm [Ce/Fe]$  &  $0.15 $  &   --                &  $ 0.10 \pm 0.08$  & $-0.04 \pm  0.14$  \\        
$ \rm [Eu/Fe]$  &  $0.02 $  &   --                &  $ 0.02 \pm 0.13$  & $-0.03 \pm  0.12$  \\                  
 \hline
\end{tabular}
}
\tablefoot{BC18: \citet{blanco18}; BC15: \citet{blanco15}; P09: \citet{pakhomov09}. Errors with $^{*}$ are lower limits.}
\end{center}
\label{tab:cr350}
\end{table}

\section{Results}
\label{res}
\begin{figure*}[h]
\centering
\hspace{-0.5cm}
\includegraphics[scale=0.9]{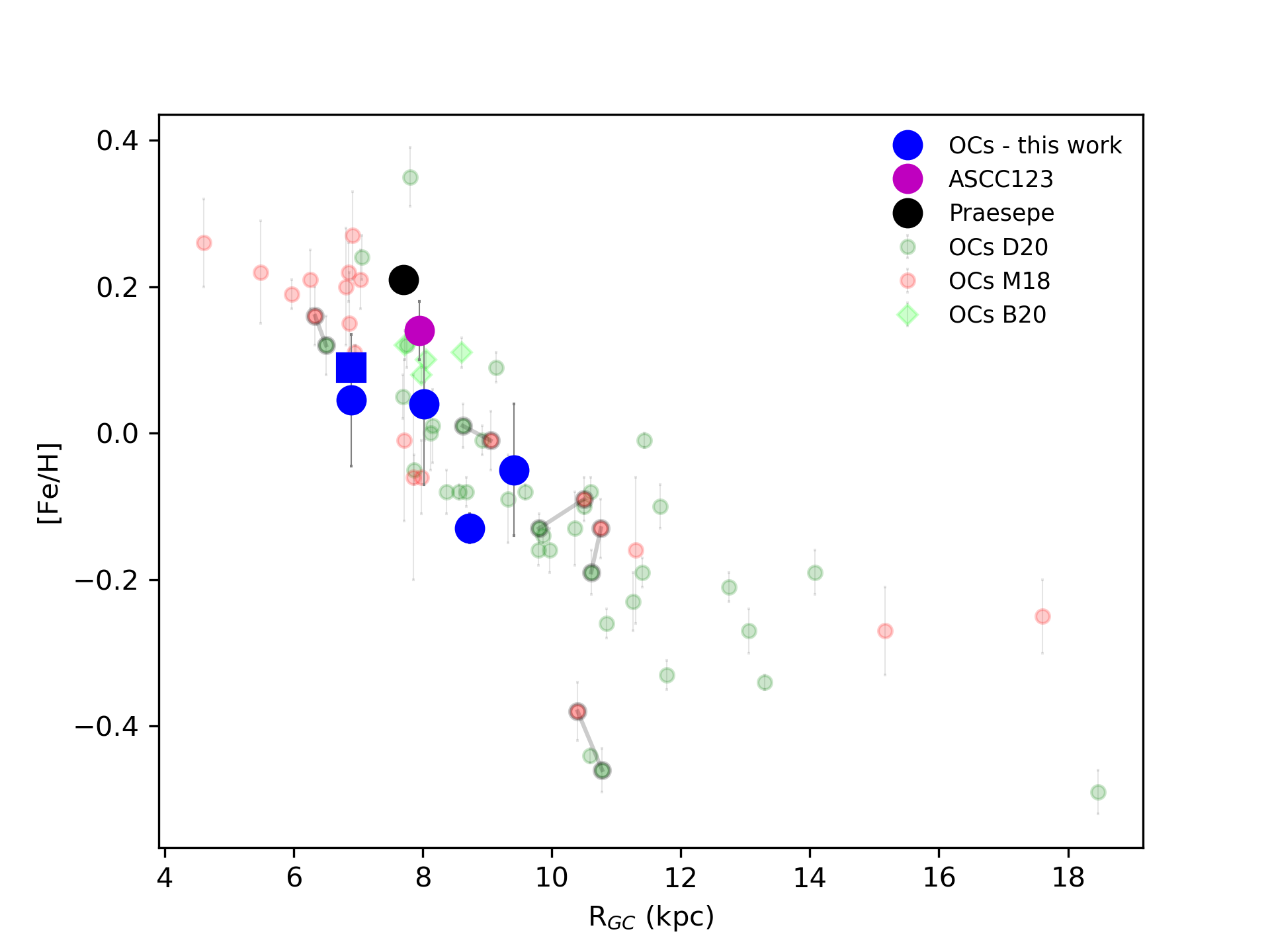} 
\caption{Radial metallicity gradient: in red the results of Gaia-ESO iDR5 from \cite{magrini18}, in green the results of APOGEE DR16 (clusters with at least three member stars and high-quality flag), and in blue our SPA clusters. The blue square is the mean [Fe/H] of Rup~171 obtained with only the hottest stars analysed by ROTFIT (ID from 3 to 8). The clusters in common between APOGEE and Gaia-ESO are linked by a black line. The magenta and black circles are the SPA clusters ASCC~123 and Praesepe studied by \citet{frasca19} and \citet{dorazi20}, respectively. The light green diamonds are the young Gaia-ESO open clusters by \citet{baratella20}. \label{fig:grad} }
\end{figure*}

Open clusters are among the best tracers of the radial metallicity distribution in the Galactic disc. Recent works within the Gaia-ESO and APOGEE surveys \citep[e.g.][]{magrini18, donor20} have investigated the shape of the radial metallicity gradient using relatively large numbers of homogeneously analysed clusters. We therefore have the opportunity to compare our sample clusters with the combined APOGEE DR16 \citep[listed by][128 clusters]{donor20}, and old and young Gaia-ESO iDR5 open clusters samples listed by \citet[][22 clusters]{magrini18} and \citet[][4 clusters]{baratella20}, respectively. Regarding the APOGEE data, we consider only open clusters with at least three member stars and with a high-quality flag, and therefore we reduce their number to 38 clusters. 
The result is shown in Fig.~\ref{fig:grad}. 
The metallicity of our cluster sample is obtained as an average between the two values shown in Table~\ref{tab:ave:met}. 
The metallicities of our four clusters, as well as ASCC~123 and Praesepe located around the Solar position, agree very well with the results of Gaia-ESO and APOGEE, and confirm an intrinsic dispersion of [Fe/H] at each Galactocentric radius. 
The innermost cluster, Rup~171, shows instead a slightly lower metallicity than the other clusters located at similar Galactocentric distances. 
However, considering only the hottest member stars of Rup~171 analysed through ROTFIT (ID from 3 to 8), we obtain [Fe/H]$\sim$0.09~dex, which is in better agreement with the global radial metallicity gradient, as shown in Fig.~\ref{fig:grad}.

In Fig.~\ref{fig:xfe,feh}, we present [X/Fe] versus [Fe/H] for the element in common among our analysis, the APOGEE results in \citet{donor20}, and the Gaia-ESO results in \citet[][for Ca, Sc, V, Cr]{magrini17}, \citet[][for Zr, La, Ce, Eu]{magrini18}, and \citet[][for Mg, Al, Si, Ti, Y]{casali20}. 
We also add the abundance ratios for two young SPA clusters: Praesepe from \citet{dorazi20}, and ASCC~123 from \citet{frasca19} -- ASCC~123 abundances scaled using the solar reference of \citet{grevesse07}. For the latter, we exclude the abundance ratios displayed as overabundant or under-abundant with respect to the field stars of  Gaia-ESO DR4  in \citet{frasca19} .

Our clusters follow the main trends of the APOGEE and Gaia-ESO clusters:  open clusters are a thin disc population, and they do not reach very low metallicities: their [Fe/H] is in the range [$-$0.5,+0.4]. 
The $\alpha$ elements (Mg, Ca, Si, Ti) all  show an enhancement in [X/Fe] towards lower [Fe/H]. Their production is essentially due to core collapse supernovae \citep[see, e.g.][]{ww95}.
ASCC~123 is overabundant in [Ca/Fe] and under-abundant in [Si/Fe] with respect to other clusters. 
Aluminium and sodium are quite scattered, which is an effect of the internal mixing that causes some sodium overabundance at the surface of red giants more massive than $\sim$1.5-2.0 M$_{\odot}$. A similar effect, even if not predicted by stellar evolution models, is observed for Al \citep[cf.][]{smi16}. 
The iron-peak elements (Sc, V, Cr, Co, Ni) follow an almost flat trend, with different degrees of scattering due to the difficulty in measuring some elements, such as for example V and Co. 
Finally, we compare five neutron capture elements, four of them predominantly produced by the slow (s) process (Y, Zr, La, Ce) and one by the rapid (r) process, Eu. 
As in \citet{casali20}, [Y/Fe] shows a peak at solar metallicity, and decreases at sub- and super-solar metallicities. 
[Zr/Fe] tends to increase towards lower metallicities, a signature of an important production in massive stars at early epochs. 
For [La/Fe] and [Ce/Fe], the general trend is similar to that of Zr, with an increasing trend at low [Fe/H].  The abundance ratios [La/Fe] of our clusters has an offset with respect to the Gaia-ESO sample. It might be due to the different solar scale for La used in \citet{magrini18}: $1.00$ dex (based on the giant stars of M67) against $1.13$ dex, determined in our solar spectrum.
Finally, Eu is an almost pure r-process element produced on shorter timescales and with a behaviour similar to the $\alpha$-elements. This behaviour is confirmed by the combination of our sample with the literature one. In general, the abundance ratios of the SPA clusters follow the main trends. 

In Fig.~\ref{fig:xfe,rgc}, we show the same abundance ratios as in Fig.~\ref{fig:xfe,feh}, plotted as a function of Galactocentric distance, $R_{\rm GC}$. 
As in Fig.~\ref{fig:xfe,feh}, our open clusters follow the main trends. 
For the $\alpha$-elements, we have a slight enhancement increasing towards the Galactic outskirts.
This enhancement is an indication of the inside-out formation of the disc, in which inner regions formed at high star formation rates, thus being quickly enriched by products of SNe Ia with a consequent lower [$\alpha$/Fe]. 
The gradients of the iron-peak elements (V, Cr, Co, Ni) are  flat, with different levels of scatter, indicating the expected similarity with the [Fe/H] gradient. The [Sc/Fe] gradient resembles the [$\alpha$/Fe] gradient, indicating a contribution from core-collapse supernovae in its nucleosynthesis. 
The [Y/Fe] gradient shows a peak at solar Galactocentric distance, and a decreasing trend towards the inner disc, as already discussed in \citet{casali20}. Even the innermost SPA open clusters do not manifest any  high [Y/Fe], but follow the trend of the literature, confirming a less efficient production of Y in the inner disc and at high metallicity. 
This behaviour is in common with the other neutron-capture elements with a predominance from s-process. The elements La and Ce share the same behaviour in the inner disc, while they are more enhanced in the outskirts as a consequence of their partial production also in massive stars. 
Finally [Eu/Fe] versus $R_{\rm GC}$ shows an increase in the outer disc, which is characteristic of a double mechanism of production of this element (Van der Swaelmen et al. in prep.).

\begin{figure*}[h]
\centering
%\hspace{-0.5cm}
\includegraphics[scale=0.5]{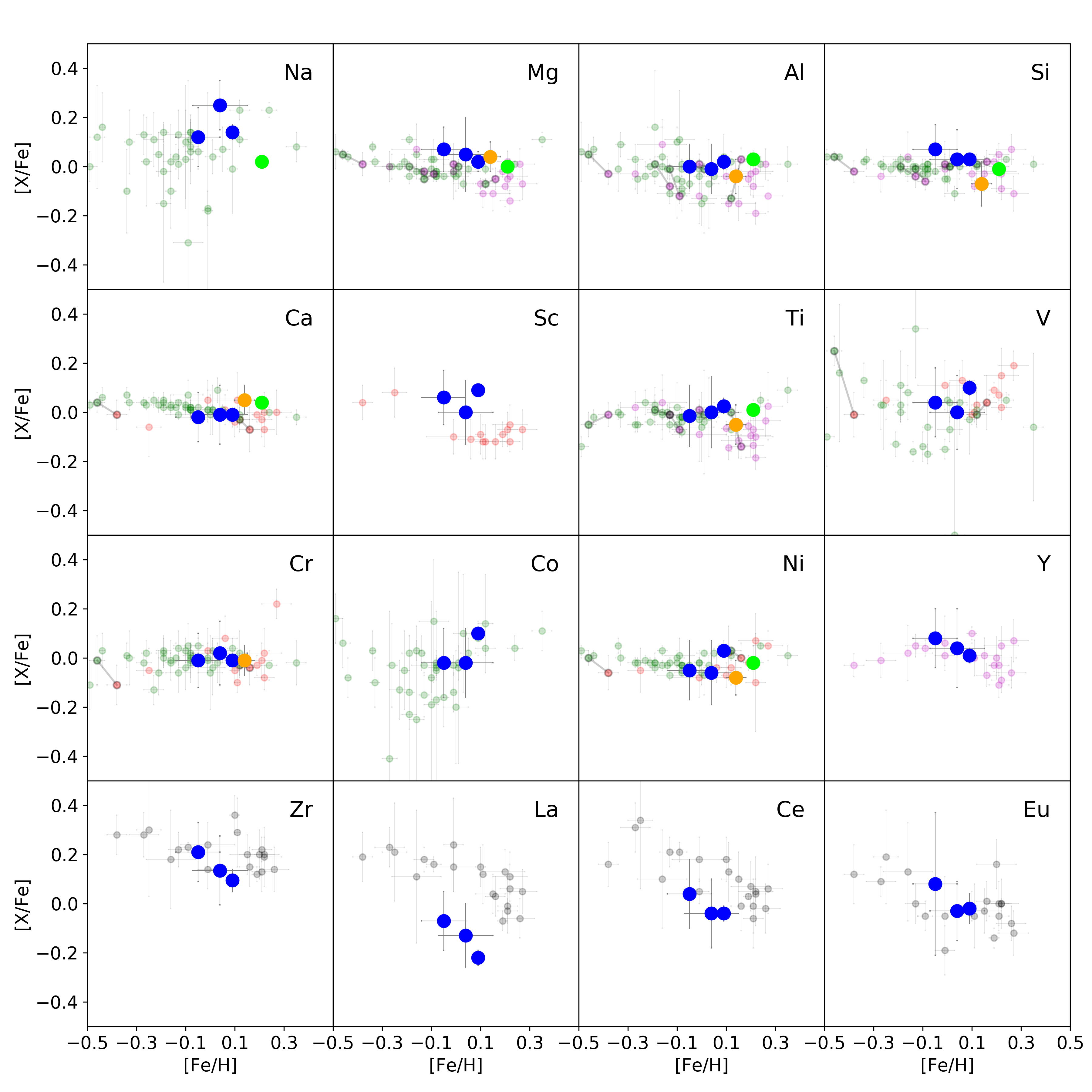} 
\caption{Abundance ratios vs. [Fe/H]: blue indicates our SPA clusters with [Fe/H] as an average between the two values shown in Table~\ref{tab:ave:met}; light green indicates Praesepe; and orange indicates ASCC~123. The transparent colours indicate the following: green circles are APOGEE DR16 open clusters listed in \citet{donor20}; red circles are the open clusters of Gaia-ESO DR4 with the [X/Fe] ratios calculated in \citet{magrini17}; magenta circles are the open clusters of Gaia-ESO iDR5 with the [X/Fe] ratios calculated in \citet{casali20}, not present in \citet{magrini17}; and finally, black circles are the open clusters of Gaia-ESO iDR5 with the [X/Fe] ratios calculated in \citet{magrini18}, not present in \citet{magrini17} and \citet{casali20}. The open clusters in common between APOGEE and Gaia-ESO are linked by a black line.\label{fig:xfe,feh} }
\end{figure*}

\begin{figure*}[h]
\centering
%\hspace{-0.5cm}
\includegraphics[scale=0.5]{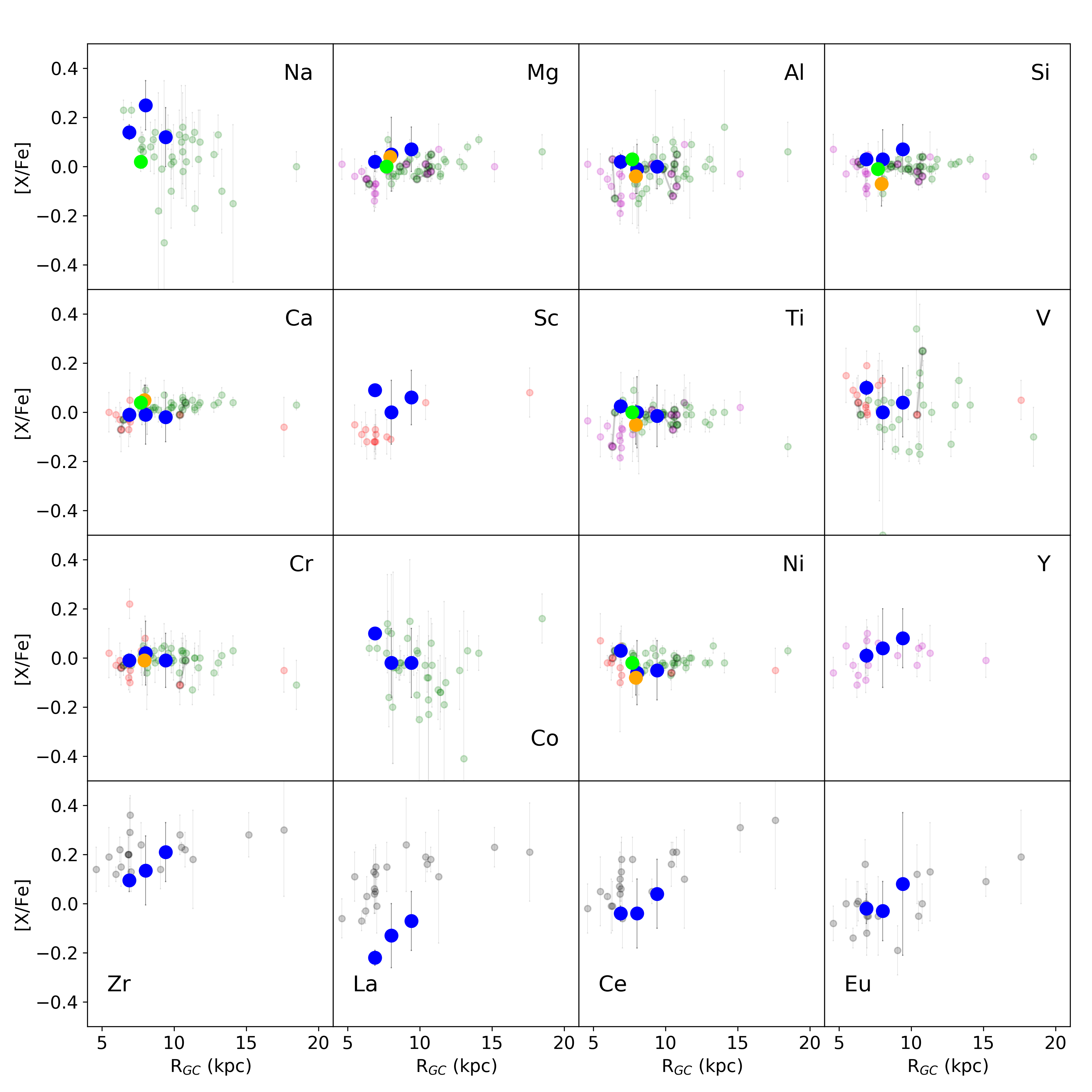} 
\caption{Abundance ratios vs. Galactocentric distance $R_{\rm GC}$. Symbols as in Fig.~\ref{fig:xfe,feh}. \label{fig:xfe,rgc} }
\end{figure*}

In Fig.~\ref{fig:afe,feh}, we present [$\alpha$/Fe] versus [Fe/H] (where [$\alpha$/Fe] is  the sum of [Mg/Fe], [Ca/Fe], [Si/Fe] and [Ti/Fe] divided by four) for our cluster sample as well as for ASCC~123 and Praesepe, together with the Gaia-ESO and APOGEE clusters, and 
the field stars from  Gaia-ESO DR4. 
The cluster population is colour coded by age using 790 Myr for Praesepe \citep{bh15,dorazi20}, the mean value of the range of 100--250 Myr suggested in \citet{frasca19} for ASCC~123, and age determinations from Table~\ref{tab:infoclusters} for our SPA clusters from \citet{magrini17} for Gaia-ESO and from \citet{donor20} for APOGEE. 
While the field population is well separated in the two components of thin and thick discs, open clusters are essentially a thin-disc population: 
most clusters are located in the low-$\alpha$ thin disc with ages lower than 5 Gyr and the SPA clusters are in agreement with the thin-disc field stars within their errors.
There are a few exceptions: the three clusters with high [Fe/H] are also slightly enhanced in [$\alpha$/Fe].  As discussed for NGC~6705 in \citet{magrini14, magrini15} and \citet{casamiquela18}, they might be part of a young metal-rich and $\alpha$-enhanced population similar to the ones found in the disc \citep[see, e.g.][]{chiappini15}, and very recently also in the bulge \citep{Thorsbro20}. 
Also in this case, the SPA open clusters follow the main trend of the thin-disc population.

\begin{figure*}[h]
\centering
\hspace{-0.5cm}
\includegraphics[scale=1]{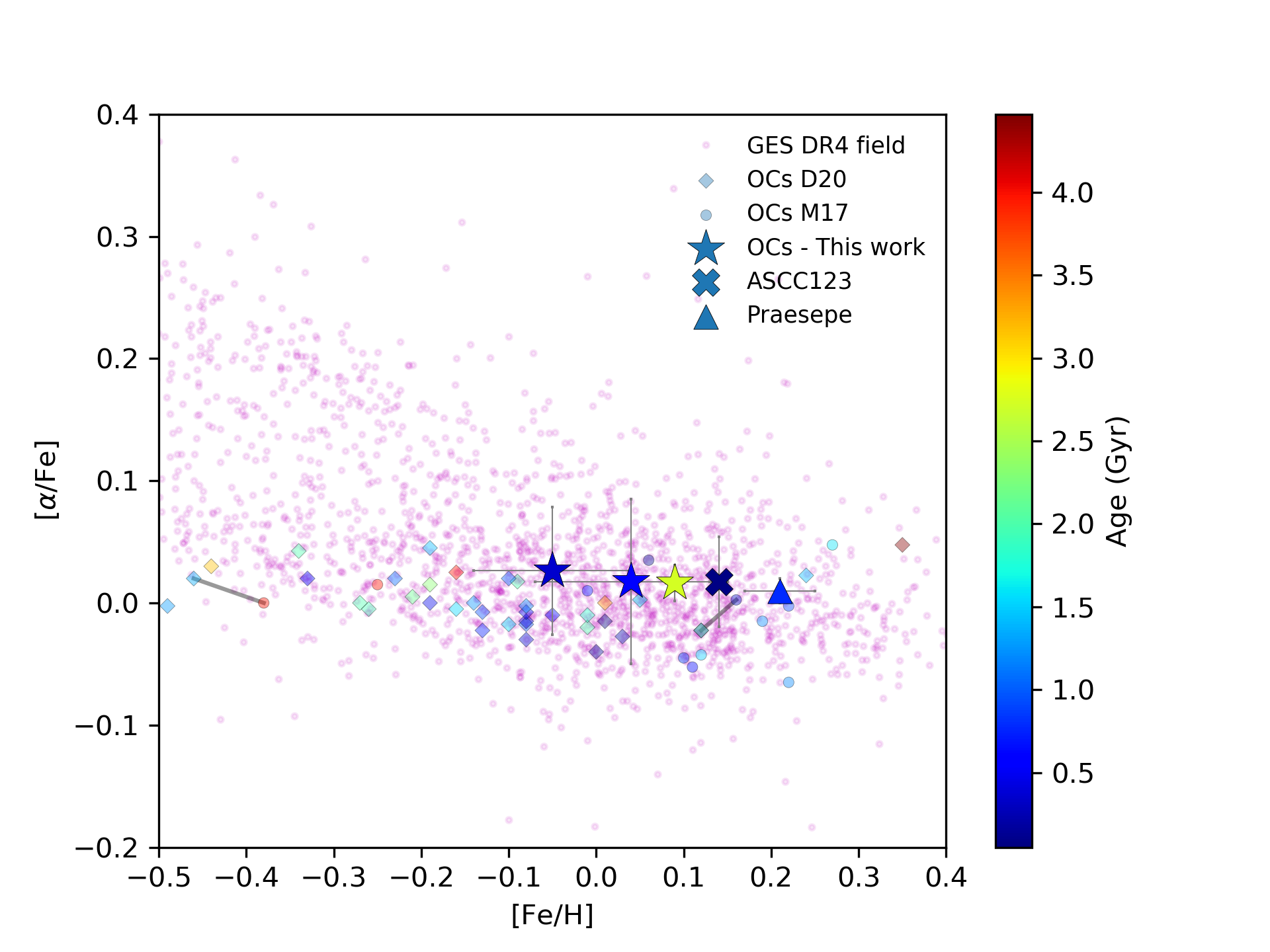} 
\caption{[$\alpha$/Fe] vs. [Fe/H] for the open clusters, colour-coded by stellar age. The SPA clusters of this work are marked with a star symbol, while the cross and the triangle are the SPA clusters ASCC~123 and Praesepe, respectively; the transparent circles are the Gaia-ESO clusters with the abundances from \citet{magrini17} and the transparent diamonds are the APOGEE clusters with the abundances from \citet{donor20}. The magenta dots (not colour-coded by age) in transparency are the field stars of the Gaia-ESO survey (public release, available at ESO archive). Finally, clusters in common between APOGEE and Gaia-ESO are linked by a black line. \label{fig:afe,feh} }
\end{figure*}

\subsection{Carbon and lithium}

We measure C and Li in some stars of our sample. 
Carbon is measured from the atomic lines \citep[the adopted C\,{\sc i} at 5052.144 $\AA$ and 5380.325 $\AA$ are not affected by strong NLTE; see e.g.][]{franchini20}), while lithium is measured from the EWs of the resonance doublet at 6708 $\AA$, which is unblended at the high-spectral resolution of HARPS-N.
The photospheric abundances of these elements are affected by stellar evolution during the RGB phase because of the first dredge up (FDU). During the FDU, the stellar convective envelope penetrates into the inner regions and brings previously processed materials to the surface, enriching the external layers in N and He, and diluting the  Li and C abundances \citep[see][for more details]{lagarde12, masseron15,salaris15,casali19}. After the FDU, the star evolves along the RGB where extra mixing such as that caused by the thermohaline mechanism likely dominates the abundance change of these elements \citep{lagarde12}. These effects are mainly expected along the upper RGB, after the RGB bump phase. The incidence of thermohaline mixing dominates at low metallicity and for low-mass stars, while it is weaker in the Solar metallicity regime. Observational evidence of such extra-mixing processes are the very low lithium abundance after the RGB bump and the variation of the isotopic ratio $^{12}$C/$^{13}$C  \citep[e.g.][]{lind09,mucciarelli11, lagarde18}.

\begin{figure*}[h]
\centering
\includegraphics[scale=0.6]{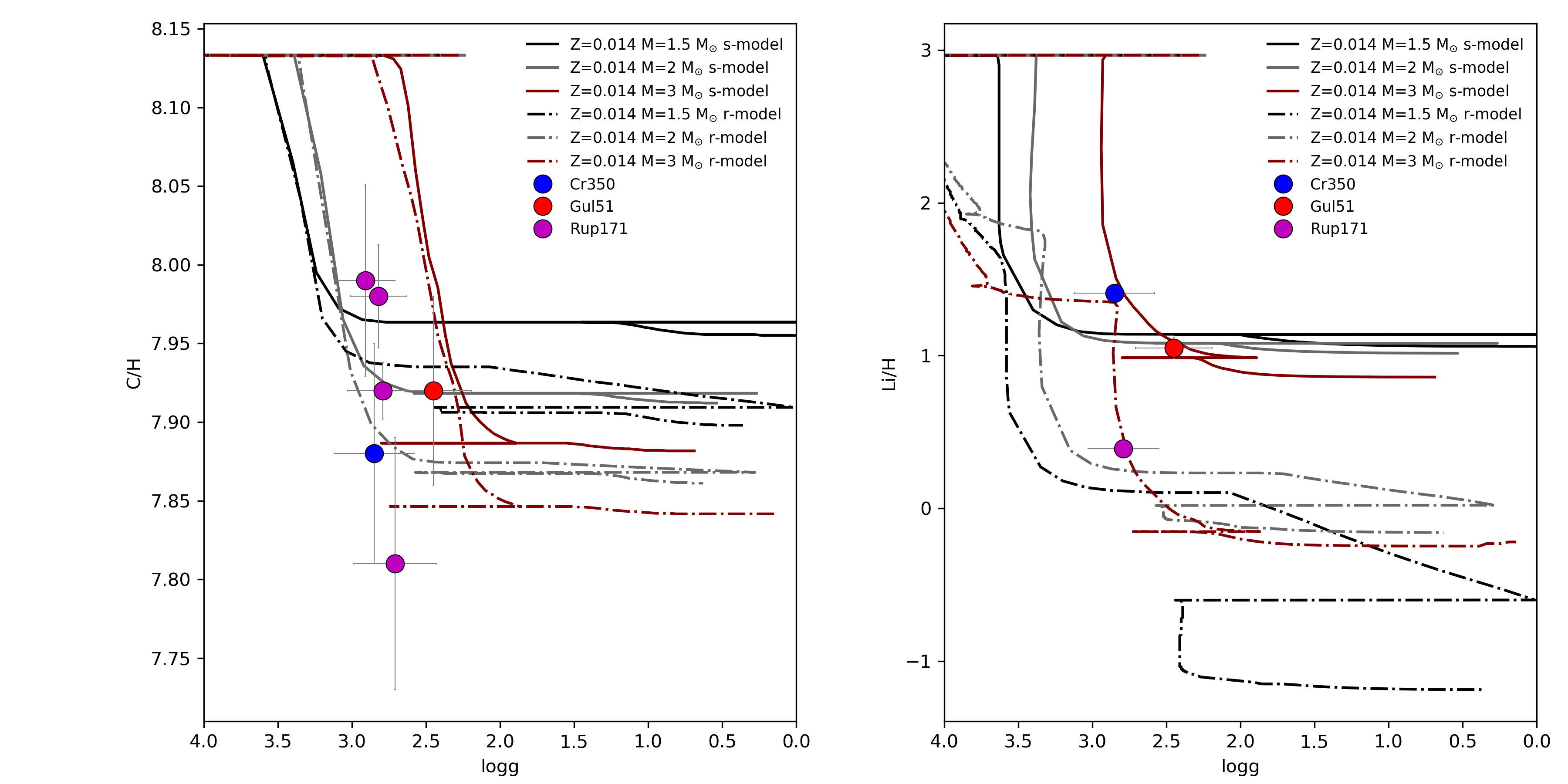} \caption{C/H and Li/H versus $\log~g$. The circles are the stars of our sample and the curves are the theoretical models by \citet{lagarde12}. Continuous lines are the standard models with masses of 1.5-2-3~M$_{\odot}$, and dashed lines are the models with rotation-induced extra-mixing for the same masses. \label{fig:c_li} }
\end{figure*}

In Fig.~\ref{fig:c_li}, we show C/H and Li/H for the stars of Cr~350, Gul~51, and Rup~171 as a function of their $\log~g$. The curves are the models of \citet{lagarde12} for three different masses (1.5, 2, 3 M$_{\odot}$, which encompass the range of turn-off masses for our clusters) for the standard (solid lines, s-models) and rotation-induced mixing models (dashed lines, r-models). The value C/H decreases with decreasing $\log~g$ for the stars in Rup~171, following the theoretical models of \citet{lagarde12} (see left panel of Fig.~\ref{fig:c_li}).  Rup171\_5 is located out of the locus of the theoretical models, however its C/H determination exhibits the largest uncertainties.
The right panel shows the evolution of Li versus $\log~g$. The stars of Rup~171, the oldest cluster of our sample, with a MSTO mass of $\sim$1.5 M$_{\odot}$, are in better agreement with models with rotation-induced mixing, while the stars in Cr~350 and Gul~51, the youngest clusters of our sample, with the highest MSTO masses ($\sim$3~M$_{\odot}$), are in agreement with the predictions of the standard models, as expected for more massive stars at Solar metallicity.

\section{Summary}
\label{conclusions}
The SPA survey is extremely useful to characterise the Solar neighbourhood, in particular the nearby clusters, providing a wide variety of elements thanks to the high spectral
resolution and large spectral coverage. 
In the framework of SPA, we present the analysis of  four poorly studied clusters: Collinder~350, Gulliver~51, NGC~7044, and Ruprecht~171. 
We analyse the high-resolution HARPS-N spectra taken at the TNG for, respectively, two, one, four, and seven candidate member stars. These stars belong to the RGB or RC of the cluster evolutionary sequence. 
We perform, for the first time for these clusters (except for Cr~350), a spectral analysis based on the EW measurements and spectral fitting to chemically characterise these clusters. 
With the EW analysis, we find a correlation between stellar parameters and metallicity for stars belonging to the same cluster, 
but in different evolutionary phases.
In particular, the coolest stars (T$_{\rm eff} < 4300$ K and $\log~g$ < 1.8 dex) appear more metal-poor than the hottest ones of the same cluster. We investigate several possibilities that might explain this phenomenon, such as the influence of photometric parameters, NLTE effects, and the use of different line lists and model atmospheres. We conclude that the continuum placement is extremely challenging for these stars, and might lead to the derivation  of incorrect metallicities. This is combined with the known inaccuracy of model atmospheres to reproduce some features of cool giants \citep{bessel98}.
On the other hand, ROTFIT provides results which do not  strongly depend on the evolutionary phase. For this reason we adopt them for the coolest stars of our sample.  
We derive chemical abundances for several elements for the stars with T$_{\rm eff} > 4300$ K. 
We investigate the Galactic radial metallicity gradient comparing our SPA clusters with those of Gaia-ESO iDR5 \citep{magrini18} and APOGEE DR16 \citep{donor20}. Collinder~350, Gulliver~51, NGC7044, and Ruprecht~171 closely follow the global radial metallicity gradient. 
We also present [X/Fe] versus [Fe/H] for elements in common with the SPA, APOGEE, and Gaia-ESO samples. The SPA open clusters follow the trends shown by the other clusters. For instance, the $\alpha$-elements (Mg, Ca, Si, Ti) show an enhancement in [X/Fe] towards lower metallicity; the iron-peak elements (Sc, V, Cr, Co, Ni) follow an almost flat trend; and the s-process elements and Eu follow the general behaviour relatively well.  
We show the [X/Fe] ratios as a function of the Galactocentric distance and the SPA clusters follow the main trends. $\alpha$-elements increase slightly towards the Galactic outskirts, implying an inside-out formation of the disc. 

We  also derive the abundances of C and Li, which are compared with the models of \citet{lagarde12}. Models with rotation-induced mixing are necessary to explain the Li abundances for the stars in the older cluster, Rup~171. No Li-rich stars are detected in our sample. 

Further developments are expected from the homogenous analysis of the whole sample of SPA clusters observed in the forthcoming runs. In addition, we plan to analyse  the GIANO spectra, both in combination with HARPS-N \citep[as done in Praesepe by][]{dorazi20} and alone, concentrating on elements that are inaccessible in the optical spectra, such as fluorine.

 \begin{acknowledgements} 
We thank Giuseppe Bono for his useful comments. This work exploits the Simbad, Vizier, and NASA-ADS
databases. We thank the TNG personnel for help during the observations. This work has made use of data from the European Space Agency (ESA) mission Gaia (https://www.cosmos.esa.int/gaia), processed by
the Gaia Data Processing and Analysis Consortium (DPAC,
https://www.cosmos.esa.int/web/gaia/dpac/consortium). Funding
for the DPAC has been provided by national institutions, in particular the institutions participating in the Gaia Multilateral Agreement.
We made use of data from the Gaia-ESO Survey Data Archive (version 2),
prepared and hosted by ESO at eso.org/qi/catalogQuery/index/121. 
LM acknowledge the funding from the INAF PRIN-SKA 2017 program 1.05.01.88.04. 
We acknowledge the funding from MIUR Premiale 2016: MITiC. 
X.F. acknowledges the support of China Postdoctoral Science Foundation No. 2020M670023, the National Natural Science Foundation of China under grant number 11973001, and the National Key R\&D Program of China No. 2019YFA0405504.
 \end{acknowledgements}

%--------------------------------------%
%-------------BIBLIOGRAPHY-------------%
%--------------------------------------%

\bibliographystyle{aa}
\bibliography{Bibliography}

\end{document}